\documentclass[aps,preprint]{revtex4}

\usepackage{amsmath}
\usepackage[caption=false]{subfig} 
\usepackage{graphicx}
\usepackage{xcolor}
\usepackage[colorlinks=true,allcolors=blue]{hyperref}
\usepackage{soul}

\begin{document}
\title{Critical fluid dynamics in two and three dimensions}

\author{Chandrodoy Chattopadhyay$^1$, 
Josh Ott$^1$, 
Thomas Sch\"afer$^1$,  
and Vladimir V. Skokov$^{1}$}
\affiliation{$^1$ Department of Physics, 
North Carolina State University,
Raleigh, NC 27695}

\begin{abstract}
 We describe a numerical method for simulating stochastic
fluid dynamics near a critical point in the Ising universality
class. This theory is known as model H, and is expected to govern
the non-equilibrium dynamics of Quantum Chromodynamics (QCD)
near a possible critical endpoint of the phase transition 
between a hadron liquid and the quark-gluon plasma. The numerical
algorithm is based on a Metropolis scheme, and automatically ensures
that the distribution function of the hydrodynamic variables in
equilibrium is independent of the transport coefficients and only 
governed by the microscopic free energy. We verify dynamic 
scaling near the critical point of a two and three-dimensional
fluid and extract the associated critical exponent $z$. We find
$z\simeq 3$ in three dimensions, and $z\simeq 2$ for a two-dimensional
fluid. In a finite system, we observe a crossover between the mean
field value $z=4$ and the true critical exponent $z\simeq 3$ ($z
\simeq 2$ in $d=2$). This crossover is governed by the values of 
the correlation length and the renormalized shear viscosity. 
\end{abstract}
\maketitle

\section{Introduction}

  Hydrodynamics provides a remarkably effective theory of many 
strongly interacting fluids \cite{Schafer:2009dj,Jeon:2015dfa,
Romatschke:2017ejr}. This description is based on a set of 
partial differential equations that express the conservation 
laws for energy, momentum, and other conserved densities. In
the long-wavelength limit the precise form of these differential
equations can be constructed order by order as an expansion in
derivatives of local thermodynamic variables. 

  It has been understood for some time that this description is
not complete and that fluctuations in the conserved densities
and currents cannot be ignored \cite{landau:SMII,Kovtun:2011np}. 
Indeed, stochastic terms in the conserved currents are constrained
by fluctuation-dissipation relations, and they cannot be zero
unless the temperature or the dissipative coefficients vanish. 
Recent work on effective actions defined on the Keldysh contour 
has demonstrated that stochastic hydrodynamic theories provide 
a systematically improvable framework for the calculation of 
correlation functions in the long-wavelength limit 
\cite{Crossley:2015evo,Chen-Lin:2018kfl,Jain:2020zhu,
Delacretaz:2020nit,Chao:2020kcf,Basar:2024srd}.

  At a generic point in the phase diagram, fluctuations lead 
to non-analytic terms in the low frequency, small momentum 
expansion of correlation and response functions. These 
contributions are known as long-time tails. They are very 
interesting, but in typical fluids they are not easy to observe. 
In the vicinity of a second-order phase transition, on the other
hand, fluctuations are large. In particular, stochastic terms in 
the hydrodynamic equations account for the divergent correlation
length and susceptibilities, and describe the phenomenon of 
dynamic scaling \cite{Hohenberg:1977ym}.

  Stochastic hydrodynamic theories were classified in the work by 
Hohenberg and Halperin \cite{Hohenberg:1977ym}. This classification
includes purely relaxational theories (model A), critical diffusion
(model B), critical anti-ferromagnets (model G), and critical 
diffusion coupled to the Navier-Stokes equation (model H). The 
last two theories are expected to be relevant to QCD. Model G 
accounts for the dynamics of the chiral phase transition
\cite{Rajagopal:1992qz}, and model H describes the dynamics 
near the endpoint of the nuclear liquid-gas transition, as well 
as a possible critical endpoint of the quark-gluon plasma 
(QGP) phase transition \cite{Son:2004iv}.

 The dynamics near a critical endpoint is genuinely non-perturbative
 -- there are no small parameters. Historically, the main approach
to the calculation of dynamical critical exponents has been the 
$\epsilon$-expansion. For the models considered here, this is an
expansion in $\epsilon=4-d$, where $d$ is the number of spatial 
dimensions \cite{Hohenberg:1977ym}. This parameter is not small, but
comparison of the results with experimental data on liquids is
encouraging \cite{Arcovito:1969}. More recently, a number of groups 
have used the functional renormalization group \cite{Canet:2006xu,
Canet:2011wf,Mesterhazy:2013naa,Chen:2023tqc,Roth:2023wbp,Chen:2024lzz,
Roth:2024hcu}. For three-dimensional systems, the results are generally 
consistent with predictions obtained in the $\epsilon$-expansion. 

These methods have some disadvantages. Up to this point, both the 
$\epsilon$-expansion and the functional renormalization group 
calculations have been performed at a rather low order of the 
corresponding truncations. More importantly, it is not obvious 
how to apply these methods to problems such as heavy-ion collisions 
that involve a non-trivial background flow. There are a number of 
recent studies that address critical dynamics in expanding fluids. 
A typical approach is to consider a non-critical fluid and derive 
a set of deterministic equations for the equal-time $n$-point 
functions of the theory \cite{Kapusta:2011gt,Mukherjee:2016kyu,
Akamatsu:2016llw,Stephanov:2017ghc,Akamatsu:2018vjr,Martinez:2018wia,
An:2019osr,An:2019csj,An:2020vri}. Critical effects can be taken
into account by replacing the susceptibilities and relaxation 
rates with their critical counterparts. 

  Despite all these efforts, there have been remarkably few 
attempts at direct simulation of stochastic hydrodynamic theories.
This is not necessarily because of a lack of interest, but because
numerical simulations encounter several problems. This includes 
the fact that short-distance noise requires regularization and
renormalization, the importance of implementing a scheme that 
respects fluctuation-dissipation relations, and the need to resolve
ambiguities in the definition of stochastic partial differential
equations. 

  Previous numerical work on stochastic fluid dynamics in a 
regime in which fluctuations are approximately described by the
linearized theory includes Refs.~\cite{Bell:2007,Donev:2010,
Camley:2010,Balboa:2012,Young:2014pka}. Classical statistical 
simulations are described in \cite{Berges:2009jz,Schweitzer:2020noq,
Schweitzer:2021iqk}. There have also been numerical studies 
of one-dimensional stochastic diffusion in a system that with 
longitudinal expansion that traverses the critical regime 
\cite{Nahrgang:2018afz,Pihan:2022xcl}. Finally, it is possible 
to use molecular dynamics with a classical Hamiltonian (for 
example a Lennard-Jones potential) that describes a fluid with 
a liquid and gas phase \cite{Kuznietsov:2022pcn,Chen:1995}. 
Simulations in the vicinity of the critical end point have been 
used to verify the critical divergence of the thermal conductivity
predicted by the $\epsilon$-expansion \cite{Chen:1995}. However,
while stochastic fluid dynamics emerges from the microscopic
simulation near the critical point, it is difficult to 
implement an arbitrary equation of state, or a general set of 
transport coefficients.

     A new approach to simulating stochastic fluid dynamics based 
on a Metropolis method is discussed in Refs.~\cite{Florio:2021jlx,
Schaefer:2022bfm,Florio:2023kmy,Chattopadhyay:2023jfm,Basar:2024qxd,
Chattopadhyay:2024jlh}. This method has been utilized in the 
mathematical literature on stochastic partial differential equations 
\cite{Gao:2021}, but it does not appear to be widely used. The idea 
is to combine the diffusive and stochastic terms in the equation of 
motion into a single Metropolis step. The advantage of this method 
is that fluctuation-dissipation relations are automatically satisfied,
and that the equilibrium distribution of the hydrodynamic variables 
is guaranteed to be controlled by the microscopic free energy 
functional. In a companion paper \cite{Chattopadhyay:2024jlh} we 
have presented the first numerical calculation of the relaxation 
rate and the dynamic critical exponent in model H. In the present
work, we provide a detailed description of the algorithm and
numerical details, consistency checks, and further applications 
to two-dimensional fluids.

\section{Stochastic fluid dynamics}
\label{sec:stoch-fl}

\subsection{Definition of Model H}
\label{sec:modH}

  We consider the long-time, fluid dynamic description of 
a conserved order parameter density $\phi$ interacting with 
the momentum density $\vec\pi$ of the fluid. This theory is 
known as Model H in the classification of Hohenberg and 
Halperin \cite{Hohenberg:1977ym,Folk:2006ve}. The equations
of motion are 
\begin{align}
\label{modH_1}
\partial_t\phi &=   \Gamma\,\nabla^2 
        \left(\frac{\delta{\cal H}}{\delta \phi}\right)
- \left(\nabla_i\phi\right) \frac{\delta{\cal H}}{\delta \pi_i^T}
      + \zeta , \\
\label{modH_2}
\partial_t \pi^T_i &= \eta \nabla^2 
    \left(\frac{\delta{\cal H}}{\delta \pi^T_i}\right)
    +  P^T_{ij} \left[\left(\nabla_j\phi\right) \frac{\delta{\cal H}}
    {\delta\phi}  \right] 
    -  P^T_{ij} \left[ 
        \nabla_k\left( \pi^{T}_j \frac{\delta{\cal H}}{\delta
       \pi^T_k}\right) \right] + \xi_i\, . 
\end{align}
Here, $\Gamma$ and $\eta$ are transport coefficients. In the 
context of describing the dynamics near a liquid-gas endpoint 
we can take $\phi$ to be the specific entropy $s/n$ of the fluid
\cite{Akamatsu:2018vjr,An:2022jgc}, $\Gamma$ is the thermal
diffusivity, and $\eta$ is the shear viscosity. The 
transverse projection operator is given by 
\begin{align}
    P^T_{ij} = \delta_{ij}- \frac{\nabla_i\nabla_j}{\nabla^2} 
\end{align}
and $\pi^T_i=P^T_{ij}\pi_j$. The effective Hamiltonian is 
given by 
\begin{align}
\label{H_Ising}
    {\cal H}  = \int d^dx \left[ 
      \frac{1}{2\rho} ( \pi_i^T)^2
    +  \frac{1}{2} (\nabla \phi)^2  
    +  \frac{1}{2} m^2 \phi^2
    +   \frac{1}{4} \lambda  \phi^4  
    - h \phi\right] \, ,
\end{align}
where $\rho$ is the mass density, $m$ is the inverse
correlation length, $\lambda$ is a non-linear self-coupling, 
and $h$ is an external field. In order to describe a possible
critical endpoint in the QCD phase diagram we identify $\rho$
with the enthalpy density of the fluid, and we map the
parameters $m^2$, $h$, and $\lambda$ onto the chemical
potential-temperature $(\mu,T)$ plane of QCD, see for 
example \cite{Parotto:2018pwx,Kahangirwe:2024cny}. The noise 
terms $\zeta$ and $\xi_i$ are random fields constrained by
fluctuation-dissipation relations. The correlation functions 
of the noise are given by  
\begin{align}
    \langle \zeta (t, \vec{x}) \zeta (t', \vec{x}') \rangle &= 
    -2 T\, \Gamma\, \nabla^2 \delta(\vec{x}-\vec{x}')\delta(t-t')\, ,
\label{noise-phi}  \\  
  \langle \xi_i (t, \vec{x}) \xi_j (t', \vec{x}') \rangle &= 
    -2 T\, \eta\, P^T_{ij} \nabla^2
      \delta(\vec{x}-\vec{x}')\delta(t-t')\, . 
\label{noise-pi} 
\end{align}

\subsection{Shear vs sound}
\label{sec:sound}

 The hydrodynamic equations (\ref{modH_1},\ref{modH_2}) describe 
the interaction of shear modes with the order parameter, but they 
do not include sound modes. This is based on the observation
that in order to study critical dynamics we can focus on the 
long-time behavior of the fluid, governed by modes with the
lowest frequency. Shear modes have a dispersion relation $\omega
\simeq i(\eta/\rho)k^2$, whereas sound modes are described by
$\omega\simeq c_sk$, where $c_s$ is the speed of sound. This means
that at long distances (or low wave numbers) the magnitude of the 
frequency of shear waves is smaller than that of sound waves. 

There are, however, phenomena that require the inclusion of sound 
modes. One of them is the critical behavior of the bulk viscosity, 
which is given by $\zeta\sim\xi^{z-\alpha/\nu}$, where $\xi$ is the
correlation length, $z\simeq 3.05$ is the dynamic exponent, $\alpha
\simeq 0.11$ is the specific heat exponent, $\nu\simeq 0.63$ is the
correlation length exponent \cite{Onuki:1997,Martinez:2019bsn}. 
Another is the dispersion relation of sound itself. Near the critical
point the speed of sound vanishes as $c_s^2\sim \xi^{-\alpha/\nu}$, 
but finite frequency modes propagate with a finite speed, governed 
by a universal function \cite{Onuki:2002,Vasilev:2004}.

Projecting on the transverse component of the momentum density
is useful, because the longitudinal part of $\vec\pi$ couples 
to the energy density ${\cal E}$ and pressure $P$ of the fluid.
This means that if $\vec\pi_L$ is included we also have to include
an equation of motion for the energy density, and we have to 
study the renormalization of the equation of state by short 
distance fluctuations. We also observe that the projection
on transverse modes simplifies the mode coupling between
$\vec{\pi}$ and $\phi$. In particular, $(\nabla_j\phi)V'(\phi)
=\nabla_jV(\phi)$ is longitudinal, so that only the gradient 
term in ${\cal H}(\phi)$ contributes to the mode-coupling between 
$\vec{\pi}$ and $\phi$.

Different authors define model H in slightly different ways. 
For example, Hohenberg and Halperin \cite{Hohenberg:1977ym} only 
consider the two mode-coupling terms between $\phi$ and $\vec\pi$, 
but not the mode-coupling of $\vec\pi$ with itself. Obviously, we 
need the first type of coupling, because otherwise there is no
interaction between the two types of fluctuations. Also, we 
have to include both the coupling of $\vec\pi^T$ to $\phi$ and 
its reverse in order to ensure that the system can equilibrate. 
The argument for neglecting the self-coupling of $\vec\pi^T$
is based on the observation that fluctuations of $\vec\pi^T$
are damped as $(\eta/\rho)k^2$, whereas critical fluctuations of 
$\phi$ are much more weakly damped, proportional to $\Gamma k^4$.
Below, we will study the role of the $\vec\pi^T$ self-coupling
numerically. We will refer to the theory defined by 
Eqs.~(\ref{modH_1},\ref{modH_2}) as model H, and the theory 
without the self-coupling as model H0. A complete model H, 
with sound modes taken into account, was studied in
Refs.~\cite{Antonov:1984,Folk:1998}, see also \cite{Onuki:2002,
Vasilev:2004}.

\subsection{Conservation laws}
\label{sec:cons}

 Two of the mode-coupling terms have a straightforward interpretation 
as  the advection terms. We can write 
\begin{align}
\partial_t\phi+\frac{1}{\rho}(\pi^T_i\nabla_i) \phi &=
      \Gamma\nabla^2  \frac{\delta{\cal H}}{\delta \phi}\,,  \\
\partial_t\pi_i^T + \frac{1}{\rho}(\pi^{T}_j\nabla_j)
  \pi^T_i &= \eta\nabla^2 
     \frac{\delta{\cal H}}{\delta \pi^T_i} + \ldots
\end{align}
On the other hand, the coupling of $\vec\pi$ to gradients of $\phi$
does not have a simple interpretation, and it goes beyond the 
standard Navier-Stokes approximation. However, it is clearly
essential in order to guarantee the symmetry of the mode-couplings,
and to ensure that energy can be exchanged between $\phi$ and 
$\vec\pi$. This term can be understood more clearly by 
writing the equations of motion in conservative form
\begin{align}
 \partial_t\phi + \vec\nabla\cdot\vec\jmath =0\, , 
 \hspace{1.5cm}
 \partial_t\pi_{T,i} + P^T_{ij}\nabla_k \Pi_{jk} = 0 \, . 
\end{align}
Here the currents contain non-dissipative as well as 
a dissipative/stochastic parts
\begin{align}
   \jmath_i = \jmath_i^{(0)} + \jmath_i^{(1)}\, ,  \hspace{1cm}
   \Pi_{ij} = \Pi_{ij}^{(0)} + \Pi_{ij}^{(1)}\, .
\end{align}
The currents are given by 
\begin{align}
   \jmath_i^{(0)} &= \phi\, \frac{\delta {\cal H}}{\delta \pi^T_i}
    =   \frac{1}{\rho}\,\phi \pi^T_{i} \, , 
    \label{j-0}\\
   \jmath_i^{(1)} &= -\Gamma\, \nabla_i 
   \frac{\delta{\cal H}}{\delta \phi} + 
   \Theta_i
   \label{j-1}
\end{align}   
as well as 
\begin{align}
   \Pi_{ij}^{(0)} &= \frac{1}{\rho} \, \pi^T_{i}\pi^T_{j}
             + (\nabla_i\phi)(\nabla_j\phi ) \, , 
   \label{Pi-0}\\
   \Pi_{ij}^{(1)} &= -\eta \left[ 
    \nabla_i \frac{\delta{\cal H}}{\delta\pi^T_{j}}
   +\nabla_j \frac{\delta{\cal H}}{\delta\pi^T_{i}}
  \right] 
  + \Lambda_{ij},
  \label{Pi-1}
\end{align}
where $\Theta_i$ and $\Lambda_{ij}$ are delta correlated noise 
fields with variance $\langle\Theta_i\Theta_j\rangle \sim 2\Gamma
T\delta_{ij}$ and 
$\langle\Lambda_{ij}\Lambda_{kl}\rangle \sim 2\eta T
(\delta_{ik}\delta_{jl}+\delta_{il}\delta_{jk})$,
respectively. This form of 
the equation makes the physical meaning of the mode coupling 
between $\partial_t\vec\pi$ and $\nabla_i\phi$ more transparent. 
It corresponds to including the stress of $\phi$ in the stress 
tensor $\Pi_{ij}$. 

 The structure of the mode couplings can also be understood
using Poisson brackets \cite{Dzyaloshinskii:1980}. We can define 
a four-component field $\psi^a=(\phi,\pi^i)$. Then, the 
mode-couplings arise from the Poisson brackets
\begin{align}
\label{EOM_PB}
    \partial_t\psi^a = 
    \left\{{\cal H},\psi^a\right\} = 
    -\int d^3x 
    \,  \left\{\psi^a,\psi^b\right\}
    \, \frac{\delta {\cal H}}{\delta\psi^b}    
   = -\int d^3x \,
      Q^{ab}   \, \frac{\delta {\cal H}}{\delta\psi^b}  \, ,
\end{align}
where the fundamental Poisson brackets $\left\{\psi^a,\psi^b\right\}$ 
are given in \cite{Dzyaloshinskii:1980}. Equation~(\ref{EOM_PB})
immediately implies that the Hamiltonian is conserved by the 
mode-coupling terms, $\partial_t {\cal H}=\{ {\cal H,H} \}=0$.
In terms of the equation of motion for $\psi^a$, we can view the 
conservation of ${\cal H}$ as a consequence of the anti-symmetry
of $Q^{ab}$, 
\begin{align}
\label{H_cons}
    \partial_t {\cal H} 
   = \int d^3x\int d^3x' \, Q^{ab}   \, 
   \frac{\delta {\cal H}}{\delta\psi^a} 
   \frac{\delta {\cal H}}{\delta\psi^b}  = 0 \, .
\end{align}
Equation~(\ref{EOM_PB}) is also useful for deriving a Fokker-Planck
equation for the time-evolution of the probability distribution
$P[\psi^a]$. In particular, the anti-symmetry of $Q^{ab}$ implies
that the Fokker-Planck equation has a stationary solution 
$\exp(-{\cal H}[\psi]/T)$ \cite{Hohenberg:1977ym}.

\section{Numerical method}
\label{sec:num}

\subsection{Lattice formulation}
\label{sec:lat}

In order to study the theory numerically we discretize the fields 
$\phi(\vec{x})$ and $\vec\pi(\vec{x})$ on a $d$-dimensional 
lattice $\vec{x}=\vec{n}a$ with $n_i=1,\ldots,N$. We can define
forward and backward derivatives in the direction $\mu=1,2,3$ 
as 
\begin{align}
\nabla^{R}_{\mu}\phi(\vec{x}) = \frac{1}{a}
 \left[\phi(\vec{x}+\hat{\mu}a)-\phi(\vec{x}) \right],
 \hspace{0.5cm}
\nabla^{L}_{\mu}\phi(\vec{x}) = \frac{1}{a}
 \left[\phi(\vec{x})-\phi(\vec{x}-\hat{\mu}a) \right].
\end{align}
Note that in the actual implementation we set $a = 1$. The 
Laplacian is defined as $\nabla^2=\nabla^L_\mu\nabla^R_\mu
= \nabla^R_\mu\nabla^L_\mu$ where summation over $\mu$ is 
implied. Note that this lattice derivative satisfies integration
by parts 
\begin{align}
    \sum_{\vec{x}} \nabla^R_\mu\phi(\vec{x})
     \nabla^R_\mu\phi(\vec{x})
 = -\sum_{\vec{x}} \phi(\vec{x})
     \nabla^2\phi(\vec{x}), 
\end{align}
where we have assumed that the fields satisfy
periodic boundary conditions.
Also note that the operators $\nabla_\mu^{R,L}$ are not 
anti-hermitian, 
\begin{align}
 \left(\nabla^R_\mu\right)^\dagger=-\nabla^L_\mu\, , \hspace{0.7cm}
 \left(\nabla^L_\mu\right)^\dagger=-\nabla^R_\mu\, , 
\end{align} 
but their average, the symmetric (centered) difference
operator is
\begin{align}
 \nabla_\mu^c=\frac{1}{2}\left(\nabla^L_\mu+\nabla^R_\mu\right)
\, , \hspace{0.7cm} 
 (\nabla_\mu^c)^\dagger = -\nabla_\mu^c \, . 
\end{align}
We can define a centered Laplacian
\begin{align}
\label{delc-2}
\left(\nabla_\nu^c\right)^2\phi(\vec{x})
 = \frac{1}{4} \sum_\nu\, \Big\{ \phi(\vec{x} + 2 \hat{\nu}) 
   + \phi(\vec{x} - 2 \hat{\nu}) - 2 \phi(\vec{x}) \Big\},
\end{align}
We are now in a position to specify the lattice discretized
Hamiltonian as 
\begin{align}
\label{H_Ising_disc}
{\cal H}  = \sum_{\vec{x}} \left[ \frac{1}{2\rho}  
         \pi^T_\mu(\vec{x})\pi^T_\mu(\vec{x})
 + \frac{1}{2} \nabla^R_\mu\phi(\vec{x})
                \nabla^R_\mu \phi(\vec{x}) 
    +  \frac{1}{2} m^2 \phi^2(\vec{x}) 
    +   \frac{1}{4} \lambda  \phi^4(\vec{x})  
    \right] \, , 
\end{align}
where again the sum over $\mu$ is implied. Below, we will 
also consider a Hamiltonian where the gradient term is 
defined in terms of centered derivatives
\begin{align}
 {\cal H}^c_{\partial\phi}  = \sum_{\vec{x}} 
 \frac{1}{2} \nabla^c_\mu\phi(\vec{x})
                \nabla^c_\mu \phi(\vec{x})
 = -\sum_{\vec{x}} \frac{1}{2} \phi(\vec{x})
   \left(\nabla^{c}\right)^2 \phi(\vec{x})\, .
\end{align}       
We define lattice transverse fields using the centered 
derivative 
\begin{align}
\left(\nabla^c\right)^2\pi_\mu^T(\vec{x})
  =    \left(\left(\nabla^c\right)^2 \delta_{\mu\nu} 
    -\nabla^c_\mu\nabla^c_\nu \right)
    \pi_{\nu}(\vec{x})\, . 
\label{pi-T-lat}    
\end{align}
It is possible to define transverse fields using the left 
and right derivatives, but in that case one has to consider 
two types of fields, left and right projected transverse
momentum densities, $\nabla_\mu^L\pi_\mu^{T,L}=0$ and 
$\nabla_\mu^R\pi_\mu^{T,R}=0$. On a discrete lattice, these
fields are not the same. 

  We can solve for the projected fields using discrete Fourier
transforms
\begin{align}
  \tilde\phi(\vec{k}) &=  \sum_{\vec{x}}
     \phi(\vec{x})  \, e^{i \vec{k}\cdot\vec{x}},\\
  \tilde\pi_\mu(\vec{k}) &= \sum_{\vec{x}}
   \pi_\mu(\vec{x})  \, e^{i \vec{k}\cdot\vec{x}},
\end{align}
where the discrete momenta are given by $\vec{k}= 2\pi\vec{m}/L$
with $m_i=0,\ldots, N-1$. The projected fields are given by 
$\tilde\pi_\mu^T=P^T_{\mu\nu}\tilde\pi_\nu$ with 
\begin{align}
   P^T_{\mu\nu}=\delta_{\mu\nu} - 
     \frac{\tilde{k}_\mu\tilde{k}_\nu}{\tilde{k}^2} \, ,
     \hspace{1cm}
     \tilde{k}_\mu=\frac{1}{a}\sin(a k_\mu)\, . 
\label{Proj-disc-1}
\end{align}

\subsection{Advection terms using skew-symmetric derivatives}
\label{sec:skew}

 In solving the equations of motion it is desirable to maintain
as many of the conservation laws and symmetries of the continuum
formulation as possible. In Sect.~\ref{sec:cons} we showed that
the equations can be written in manifestly momentum and charge 
conserving forms. We also saw that the symplectic structure of the 
advection term implies the conservation of ${\cal H}$. It is 
instructive to see how this works in more detail. In the continuum
the advection term for $(\phi, \vec{\pi}^T)$ is 
\begin{align}
    \dot{\phi} &= - \nabla_i \left( \phi \, 
    \frac{\pi^T_i}{\rho} \right) 
    = - \frac{\pi^T_i}{\rho} \, \nabla_i \phi, 
\label{phi_dot}\\
    \dot{\pi}^T_{i} &= - P^T_{ij}\, \nabla_k 
    \left( \frac{1}{\rho}  \pi^T_{k} \pi^T_{j} 
    + \nabla_k \phi \nabla_j \phi \right) \nonumber \\
    & = - P^T_{ij} \left[ \nabla_k 
    \left( \frac{1}{\rho} \, \pi^T_{k} \pi^T_{j} \right) 
    + \nabla_j \phi \nabla^2 \phi \right]. 
\label{pi_dot}
\end{align}
The time derivative of the Hamiltonian is
\begin{align}
    \dot{{\cal H}} = \int d^3x \, \left[ - \dot{\phi} \, \nabla^2 \phi + \frac{1}{\rho} \, \pi^T_{i} \dot{\pi}^T_{i} + V'(\phi) \, \dot{\phi} \right].
\end{align}
Substituting the equations of motion (\ref{phi_dot}-\ref{pi_dot})
we obtain
\begin{align}
    \dot{{\cal H}} = \int d^3x \, \left[  (\nabla^2\phi) \, 
     \frac{\pi^T_i}{\rho}  \nabla_i \phi 
    - \frac{\pi^T_i}{\rho} \left( \frac{\pi^T_j}{\rho} \nabla_j \right) 
    \pi^T_i 
    - (\nabla^2\phi) \, \frac{\pi^T_i}{\rho} \nabla_i \phi 
    - \nabla_i \left( V(\phi) \frac{\pi^T_i}{\rho} \right)  \right]. \label{H_dot}
\end{align}
The first and third terms cancel each other. The second term can 
be written as a divergence
\begin{align}
    \frac{\pi^T_i}{\rho}  \left( \frac{\pi^T_j}{\rho} 
    \nabla_j \right) \pi_i^T = \nabla_i  
    \left( \frac{\pi_i^T}{\rho} \, \frac{\pi^T_j\pi^T_j}{2 \,\rho} 
    \right)\, , 
\label{pi_div}    
\end{align}
and the total energy is conserved as long boundary terms can be 
ignored. However, these manipulations are not necessarily 
allowed in the discretized theory. This includes the vector 
identities used in Eq.~(\ref{pi_div}), as well as the integration
by parts identities used to simplify the gradient and potential 
terms for $\phi$.

  The question to what extent vector identities in the continuum
theory can be maintained in a lattice discretized theory was studied
systematically by Morinishi et al.~\cite{Morinishi:1998}. They first 
observe that among several ways of writing the self-advection term
of the momentum density the ``skew-symmetric'' form 
\begin{align}
\left.\nabla_\mu 
 \left(\frac{1}{\rho} \pi^T_{\mu} \pi^T_{\nu} \right) \right|_{\it skew}
 \equiv \frac{1}{2} \, \nabla_\mu 
  \left( \frac{1}{\rho} \pi^T_{\mu} \pi^T_{\nu} \right) 
  + \frac{1}{2} \, \frac{\pi^T_{\mu}}{\rho} \, \nabla_\mu \pi^T_{\nu} 
\label{skew}  
\end{align}
is distinguished by the fact that conservation of kinetic energy 
holds independent of the continuity relation $\nabla_k\pi^T_k=0$.
The second observation is that Eq.~(\ref{pi_div}) can be preserved
with a suitable definition of the derivative and the product of 
two fields. Consider the lattice interpolation of a field $\phi$
\begin{align}
\overline{\phi}^{\,\hat\mu} \equiv 
  \frac{1}{2}\big[
  \phi\left(\vec{x} + \hat\mu/2 \right) 
  + \psi\left(\vec{x} - \hat\mu/2 \right) \big]\, , 
\label{lattice-int}  
\end{align}
the lattice interpolation of the product of two fields $\phi$ and 
$\psi$
\begin{align}
  (\widetilde{\phi \, \psi})^{\hat\mu} \equiv 
  \frac{1}{2}\big[
  \phi\left(\vec{x} + \hat\mu/2 \right) 
  \psi\left(\vec{x} - \hat\mu/2 \right) 
 +  \phi\left(\vec{x} - \hat\mu/2 \right) 
    \psi\left(\vec{x} + \hat\mu/2 \right)\big],
\label{lattice-prod}    
\end{align}
as well as the symmetric half-step derivative
\begin{align}
\nabla^{1/2}_\mu \phi(x) = 
 \frac{1}{2} \big[
  \phi(x+\hat{\mu}/2) - \phi(x - \hat{\mu}/2) \big]\, . 
\label{nabla-half}  
\end{align}
Then the following product rule holds
\begin{align}
\nabla^{1/2}_\mu \left( \widetilde{\phi \, \psi} \right)^{\hat{\mu}}
 = \left(\nabla^c_{\mu} \phi\right)\psi
   + \phi \, \left(\nabla^c_\mu \psi\right)\, , 
\end{align}
and we can show that 
\begin{align}
\frac{1}{2}\, \frac{d}{dt}\, \left(\pi^T_{\nu}\pi^T_\nu\right) = 
 \left.\pi^T_{\nu} \nabla_\mu
 \left(\frac{1}{\rho} \pi^T_{\mu} \pi^T_{\nu} \right) \right|_{\it skew}
 = \frac{1}{2}\, \nabla_\mu^{1/2}
  \Big[(\overline{\pi^T_{\mu}})^{\hat{\mu}} 
       (\widetilde{\pi^T_{\nu}\pi^T_{\nu}})^{\hat{\mu}}\Big]\, , 
\label{cons-pi2}       
\end{align}
which ensures that the advection step conserves the kinetic energy
of the fluid. 

 A similar construction is possible for the mutual advection of $\phi$
and $\pi$. That case is more complicated, because the equations of motion
contain third derivative terms in $\phi$. Also, in the continuum, there 
are two separate conservation laws. One conserved quantity is the sum 
$[(\pi^T_i)^2/\rho+(\nabla_i\phi)^2]/2$, the other is the potential 
energy $V(\phi)$. On the lattice we can only implement the first of
these conservation laws exactly. For this purpose we write the mutual
advection terms using centered derivatives
\begin{align}
\dot\phi=- \frac{1}{\rho}\,\pi^T_{\mu} \nabla^c_{\mu} \phi\, , 
    \hspace{0.5cm}
\dot\pi^T_\mu = - \left(\nabla_\mu^c\phi\right)
   \left(\nabla_\nu^c\nabla_\nu^c\phi\right)\, ,
\label{phi-pi-adv}
\end{align}
where the centered Laplacian is defined in Eq.~(\ref{delc-2}).
The update in Eq.~(\ref{phi-pi-adv}) has the property that it 
conserves the Hamiltonian
\begin{align}
    {\cal H}_\pi+{\cal H}^c_{\partial\phi}= 
    \sum_{\vec{x}} \left[ 
    \frac{1}{2\rho} \pi^T_\mu(\vec{x})\pi^T_\mu(\vec{x})
 + \frac{1}{2} \nabla^c_\mu\phi(\vec{x})
                \nabla^c_\mu \phi(\vec{x}) \right] \, .
\end{align}
Note that the update does not exactly conserve the potential energy, 
nor does it conserve the gradient energy of the scalar field 
computed from forward derivatives, as in Eq.~(\ref{H_Ising_disc})
\footnote{Note that we could have used the centered gradient
of $\phi$ in the dissipative step. However, this choice makes
the stochastic update more non-local, and significantly increases
the complexity of a checker board update. In practice, exactly 
conserving the centererd derivative kinetic energy significantly 
improves conservation of the forward derivative kinetic energy.}. 
We will study the effect of this on our numerical simulations in 
Section \ref{sec:num}. 

 In summary, we use the following spatial discretization of the 
advection step
\begin{align}
\dot{\phi} &= - \frac{1}{\rho}\, 
  \pi^T_{\mu} \, \nabla^c_{\mu} \phi\, ,  
\label{phi_dot_discrete} \\
 \dot{\pi}^T_{\mu} &= - \left[ \frac{1}{2} \nabla^c_{\nu} 
   \left( \frac{1}{\rho} \pi^T_\nu \pi^T_\mu \right) 
 + \frac{1}{2\rho} \pi^T_\nu \, \nabla^c_{\nu} \pi^T_\mu
    + \left(\nabla^c_\mu \phi\right) 
      \left(\nabla^c_\nu\nabla^c_\nu\phi\right)  \right]\, . 
\label{pi_dot_discrete}
\end{align}
This update does not preserve the transversality of $\pi^T_\mu$.
After each discrete time step (see next section), we apply the 
transverse projection operator in Eq.~(\ref{Proj-disc-1}), 
which ensures that $\pi^T_\mu$ is transverse with regard
to the centered derivative, $\nabla_\mu^c\pi^T_\mu(\vec{x})=0$.
Based on Eq.~(\ref{H_dot}), this implies that energy 
is conserved up to finite lattice spacing corrections.

\subsection{Third order Runge-Kutta}
\label{sec:RK3}

 We integrate the advection terms 
Eqs.~(\ref{phi_dot_discrete}-\ref{pi_dot_discrete}) using a Runge-Kutta
method. We define a four-component field $\phi_\mu = (\phi, \vec{\pi}^T)$
and write the evolution equations as
\begin{align}
    \dot{\phi}_\mu = {\cal F}_\mu (\phi_\nu)\, . 
\end{align}
We integrate this equation using the strong stable third order 
Runge-Kutta (RK3) scheme developed by Shu and Osher \cite{Shu:1988}.
The RK3 scheme is defined by
\begin{align}
\phi^{n+1/3}_\mu &= \phi^n_\mu 
  + \Delta t \, {\cal F}_\mu (\phi^n_\nu), \\
 \phi^{n+2/3}_\mu &= \frac{3}{4} \, \phi^n_\mu 
   + \frac{1}{4} \, \phi^{n+1/3}_\mu 
   + \frac{\Delta t}{4} \, {\cal F}_\mu\left( \phi^{n+1/3}_\nu \right), \\
 \phi^{n+1}_\mu &= \frac{1}{3} \, \phi^{n}_\mu 
   + \frac{2}{3} \, \phi^{n+2/3}_\mu 
   + \frac{2}{3} \, \Delta t \, {\cal F}_\mu 
             \left( \phi^{n+2/3}_\nu \right).
\end{align}
We apply the transverse projector (\ref{Proj-disc-1}) after every 
substep of the algorithm.

\subsection{Dissipative update}
\label{sec:diss}

  We perform the dissipative update using the Metropolis algorithm 
recently studied in Refs.~\cite{Florio:2021jlx,Schaefer:2022bfm,
Chattopadhyay:2023jfm,Florio:2023kmy,Basar:2024qxd}, see also recent 
work in the mathematical literature \cite{Gao:2021}. The basic 
observation is that the diffusive step and the noise term can be 
realized by a single Metropolis update. Here, the first moment of 
the Metropolis step realizes the diffusive step, and the second 
moment implements the noise term. By combining the two terms we 
guarantee that fluctuation-dissipation relations are satisfied, 
and that the algorithm converges to an equilibrium distribution
which only depends on the Hamiltonian, and not on the dissipative 
coefficients. In equilibrium the probability density of the fields 
is given by the Gibbs distribution $P[\phi_\mu]\sim \exp(-{\cal H}
[\phi_\mu]/T)$.

 The Metropolis update for the field $\phi$ is identical to the 
update previously used in model B \cite{Chattopadhyay:2023jfm}. We 
have
\begin{align}
\phi^{\it trial}(\vec{x},t+\Delta t) & = \phi(\vec{x},t) 
   + q_\mu\, ,
\label{phi-stoch-1}   \\
\phi^{\it trial}(\vec{x}+\hat\mu,t+\Delta t) &= 
   \phi(\vec{x}+\hat{\mu},t)  - q_\mu\,  , 
\label{phi-stoch-2} \\
 q_\mu &= \sqrt{2\Gamma T(\Delta t)}\, \xi  \, ,
\label{phi-stoch_3}
\end{align}
where $\xi$ is a Gaussian random variable with unit variance 
and $\hat{\mu}$ is an elementary lattice vector in the direction 
$\mu=1,\ldots,d$. The update is accepted with probability 
${\it min}(1,e^{-\Delta{\cal H}/T})$. Note that this algorithm is 
automatically conserving. We follow the same procedure for $\vec\pi$ 
and perform a trial update 
\begin{align}
\pi_\nu^{\it trial}(\vec{x},t+\Delta t) & = 
         \pi_\nu(\vec{x},t) + r_{\nu}^{(\mu)}\, ,
\label{pi-stoch-1} \\
\pi_\nu^{\it trial}(\vec{x}+\hat\mu,t+\Delta t) &= 
  \pi_\nu(\vec{x}+\hat{\mu},t)  - r_{\nu}^{(\mu)}\,  ,
\label{pi-stoch-2} \\
 r_{\nu}^{(\mu)} &= \sqrt{2\eta T (\Delta t)}\, 
   \zeta_{\nu}^{(\mu)}  \, , 
\label{pi-stoch-3}
\end{align}
where $\zeta_\nu^{(\mu)}$ are Gaussian random variables with 
$\langle \zeta_\mu^{(\alpha)}\zeta_\nu^{(\beta)}\rangle=
\delta_{\mu\nu}\delta^{\alpha\beta}$. Again, the update is
accepted with probability ${\it min}(1,e^{-\Delta{\cal H}/T})$. 
After a complete sweep through the lattice we project on the 
transverse component of momentum density, $\pi_\mu^T (\vec{x},t)
=P^T_{\mu\nu} \pi_\nu(\vec{x})$. The projection is carried out 
in Fourier space, as described in Sect.~\ref{sec:lat}.
The projection ensures that there are $(d-1)$ fluctuating 
degrees of freedom for the momentum density (in $d$ 
dimensions), in accordance with Eq.~(\ref{noise-pi}). It also 
ensures that the average energy stored in a momentum density
mode with wave number $\vec{k}$ is equal to $\frac{d-1}{2}T$ 
rather than $\frac{d}{2}T$.

 For the Metropolis accept-reject step we have to compute the change 
in Hamiltonian due to the change in the fields. The change in ${\cal H}$
due a local change in $\phi$ at $\vec{x}$ is 
\begin{align}
\label{Delta_H_modA}
\Delta {\cal H}_\phi  (\vec{x})&= 
 d \left[  \left(\phi^{\it trial}(\vec{x})\right)^2 
             - \left(\phi(\vec{x})\right)^2 \right]
    -  \left(\phi^{\it trial}(\vec{x}) - \phi(\vec{x})\right)
    \sum_{\mu=1}^d \left[ \phi(\vec{x}+\hat{\mu}) 
                               + \phi(\vec{x}-\hat{\mu})  \right]
  \nonumber \\   
    &+\frac{1}{2} m^2 \left[ 
    \left(\phi^{\it trial}(\vec{x}\right)^2 
     - \left(\phi(\vec{x})\right)^2 \right]  
    +   \frac{1}{4} \lambda  \left[ 
        \left(\phi^{\it trial}(\vec{x})\right)^4 
           - \left(\phi(\vec{x})\right)^4\right]  \,.
\end{align}
Here, we use the lattice Hamiltonian defined in Eq.~(\ref{H_Ising_disc}).
A conserving update involves the transfer of charge $q_\mu$ from 
$\vec{x}+\hat{\mu}$ to $\vec{x}$. The change in ${\cal H}$ is 
\begin{align}
\label{Delta_H_modB}
\Delta {\cal H}_\phi (\vec{x}, \vec{x}+\hat \mu) &=  
    \Delta {\cal H}_\phi(\vec{x})  
    + \Delta {\cal H}_\phi (\vec{x}+\hat \mu) 
     + \left(q_\mu\right)^2\, .
\end{align}
The Hamiltonian is local and quadratic in the momentum density. 
This implies that the change in the Hamiltonian due to a transfer 
of momentum is particularly simple
\begin{align}
\label{Delta_H_modH}
    \Delta {\cal H}_\pi (\vec{x}, \vec{x}+\hat \mu) = 
     \frac{1}{\rho} \left[  r_{\nu}^{(\mu)} 
     \left( \pi^T_\nu(\vec{x})-\pi^T_\nu(\vec{x}+\hat{\mu})  \right)
       + \left(  r_{\nu}^{(\mu)} \right)^2
     \right]\, . 
\end{align}
The Metropolis updates can be performed on a checkerboard as explained 
in Ref.~\cite{Chattopadhyay:2023jfm}. Also note that as described here 
the timestep $\Delta t$ for the dissipative update is the same as the 
timestep in the Runge-Kutta update, but there is no requirement for 
the two to be the same. Finally, the update of the momentum density 
does not preserve the transversality of $\pi_\mu^T$. After a complete
sweep through the lattice we apply the transverse projection operator
defined in Eq.~(\ref{Proj-disc-1}).

\subsection{Choice of units}
\label{sec:units}

 The equations are solved on a spatial lattice with lattice constant
$a$. In the following we will set $a=1$, and all distances are measured
in units of $a$. At the critical point the theory is scale invariant, 
and the value of $a$ in physical units does not play a role. Away 
from the critical point the correlation length $\xi$ in units of $a$ 
is finite, and the value of $\xi$ in QCD (or any other microscopic 
theory) can be used to fix the value of $a$ in units of meters. 
Note that stochastic fluid dynamics is an effective 
theory, and it may not be possible or desirable to take the 
regulator $\Lambda\sim 1/a$ to infinity while keeping physical 
low energy parameters of the theory fixed. For example, the 
perturbative one-loop expression in Eq.~({\ref{eta-R}}) suggests
that the physical viscosity goes to infinity as $\Lambda\to\infty$
for any value of the bare viscosity. 

The  conductivity $\Gamma$ has units of $l^4/t$, where $t$ is time 
and $l$ is length. We will set $\Gamma=1$, and time is measured in 
units of  $a^4/\Gamma$. The scalar field has units of $(T_0/a)^{1/2}$, 
where $T_0$ is a temperature (and the Boltzmann constant is set to 
one). We can adopt the choice of units $T_0=1$ and measure $\phi$ 
in units of $(T_0/a)^{1/2}$. The momentum density $\pi$ is then 
measured in units of $T_0/\Gamma$, and the mass (or enthalpy) density 
is expressed in units of $\rho_0=T_0a^3/\Gamma^2$. Finally, the 
viscosity is measured in units of $\eta_0=T_0a/\Gamma$, and the 
specific viscosity $\eta/\rho$ in units of $\Gamma/a^2$. Note that 
we have set $\Gamma=1$ by a choice of units, but the dimensionless 
viscosity and mass density are parameters of our theory.

 The free energy density has three additional parameters, the mass $m$
which is the inverse bare correlation length, the non-linear self 
coupling $\lambda$, and an external field $h$. At the critical point 
the system is universal and we can pick a value of $\lambda$ and tune 
to the critical point $h=0$ and $m^2=m_c^2$. Away from the critical 
point $\lambda$ is a relevant parameter. 

 Let us provide a simple estimate of ``realistic'' values of the 
parameters, which correspond to a quark gluon fluid near a possible
critical end point. At the critical point the theory is scale invariant,
and the choice of $a$ is arbitrary. Away from the critical point the 
renormalization of the viscosity described in Eq.~(\ref{eta-R}) 
below implies that $a$ cannot be made arbitrarily small while 
maintaining the physical value of the shear viscosity. In practice
it does not make sense to make $a$ much smaller than the non-critical
value of the correlation length. This quantity is not very well 
constrained, but typical estimates are $\xi_0\simeq 0.75$ fm 
\cite{Martinez:2019bsn,Akamatsu:2018vjr}. Then $\Delta t=a^4/\Gamma
\simeq 0.32$ fm. Consider a critical endpoint at $T_0=130$ MeV. A very
rough estimate of the critical enthalpy density, based on non-interacting
quarks and gluons, is $w/(s_0T_0)\simeq 11.2$ where $s_0=(\Delta t)^2
/a^5$. If we take the bare viscosity to be $\eta/s=1/(4\pi)$, then 
the entropy in lattice units is given by $\eta/\eta_0=0.5$.
  
\section{Theoretical Expectations}
\label{sec:theory}

\subsection{Statics}
\label{sec:stat}

  The static behavior of the model is governed by the partition
function 
\begin{align}
    Z = \int D\phi\, D\vec\pi^T\, \exp\left(-\frac{{\cal H}}{T}\right),
\end{align}
where ${\cal H}$ is given in Eq.~(\ref{H_Ising}). There is no coupling
between $\phi$ and $\vec\pi^T$ in the Hamiltonian, and the integral over
the momentum density is Gaussian. This means that the equal time 
correlation function of $\vec\pi^T$ is given by 
\begin{align}
 \langle \pi^T_i(0,\vec{x})\pi^T_j(0,\vec{x}^{\,\prime})\rangle = 
   T\rho\, P^T_{ij}\delta^3(\vec{x}-\vec{x}^{\,\prime}).
\end{align}
This implies that the spectral density $\langle \pi^T_i(0,\vec{k})\pi^T_j
(0,-\vec{k})\rangle = P_{ij}^T\rho_\pi(\vec{k})$ is independent of 
$\vec{k}$, which is the classical equidistribution law. 

 The scalar field $\phi$ is governed by a Hamiltonian in the universality
class of the Ising model. This implies that for any value of $\lambda$
we can tune $m^2$ to a critical point $m_c^2$ at which the theory becomes
scale invariant. At the critical point the equal-time two-point
function of $\phi$ is 
\begin{align}
    \langle \phi(0,\vec{x})\phi(0,\vec{x}^{\,\prime})\rangle
     \sim |\vec{x}-\vec{x}^{\,\prime}|^{-d+2-\eta^*}\, , 
     \hspace{1cm}
     \eta^*\simeq 0.0363 \;\; (d=3)\, ,
\label{def:eta}
\end{align}
where we have used $\eta^*$ to distinguish the correlation function
exponent from the viscosity $\eta$, and we have quoted the critical 
exponent determined in Ref.~\cite{Alday:2015ota}. Away from the critical
point the correlation length is finite 
\begin{align}
    \langle \phi(0,\vec{x})\phi(0,\vec{x}^{\,\prime})\rangle
     \sim \frac{e^{-|\vec{x}-\vec{x}^{\,\prime}|/\xi}}
       {|\vec{x}-\vec{x}^{\,\prime}|}\, ,
       \hspace{0.5cm}
       \xi\sim |m^2-m_c^2|^{-\nu}, 
       \hspace{0.5cm}
       \nu\simeq 0.62999(5) \;\; (d=3),
\label{def:nu}
\end{align}     
with the value of $\nu$ taken from \cite{El-Showk:2014dwa}. The 
integral of the two-point function defines the susceptibility
\begin{align}
  \chi = \int d^dx\, \langle \phi(0,\vec{0})\phi(0,\vec{x})\rangle , 
   \hspace{0.5cm}
   \chi \sim |m^2-m_c^2|^{-\gamma},
   \hspace{0.5cm}
   \gamma=\nu(2-\eta)\, . 
\label{def:gamma}
\end{align}
Note that the small value of $\eta^*$ implies that the susceptibility
is close to the mean field prediction $\chi\sim\xi^2$, and that the 
two-point function in momentum space is well described by the 
Ornstein-Zernike form $\langle \phi(0,\vec{k})\phi(0,-\vec{k})\rangle 
\sim \xi^2/(1+(k\xi)^2)$.

\subsection{Dynamics: Momentum density}
\label{sec:dyn-eta}

\begin{figure}[t]
\subfloat[]{
\includegraphics[width=0.30\columnwidth]{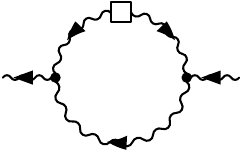}}
\hspace*{0.01\columnwidth}
\subfloat[]{
\includegraphics[width=0.30\columnwidth]{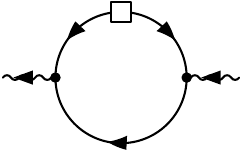}}
\hspace*{0.01\columnwidth}
\subfloat[]{
\includegraphics[width=0.30\columnwidth]{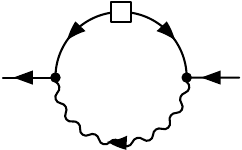}}\\
\subfloat[]{
\includegraphics[width=0.30\columnwidth]{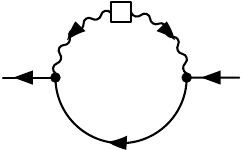}}
\caption{Loop corrections to the retarded correlation function 
of hydrodynamic fluctuations. (a) Renormalization of the momentum
correlation function by self-advection. (b) Renormalization of 
$C_\pi$ due to the coupling between $\pi$ and $\phi$. (c) and (d)  
Loop correction to the correlation function of $\phi$ due 
to advection by $\pi$.}
\label{fig:diags}
\end{figure}

 The main new degree of freedom in model H is the momentum density 
$\pi^T_i$ of the fluid. This dynamics of $\pi_i^T$ can be accessed by
studying the correlation function 
\begin{align}
 C_{ij}(t,\vec{k}) = \left\langle \pi_i^T(0,\vec{k})
    \pi_j^T(t,-\vec{k}) \right\rangle \, .
\label{C-pi}
\end{align}  
The transversality of $\pi_i^T$ implies that $C_{ij}(t,\vec{k})= (\delta_{ij}
-\hat{k}_i\hat{k}_j) C_\pi(t,k)$. In linearized hydrodynamics this correlation
function is governed by the shear mode and \cite{landau:SMII}
\begin{align}
    C_\pi(t,k) = \rho T\, \exp\left(-(\eta/\rho)k^2t\right)\, .
\label{C-pi-lin}
\end{align}
Non-linear effects will lead to a number of modifications of this 
result, even away from the critical point. The first is a 
renormalization of the viscosity due to hydrodynamic fluctuations 
and the self-advection of $\pi^T_i$, see Fig.~\ref{fig:diags}(a). 
Here, we follow Ref.~\cite{Chao:2020kcf} and use a diagrammatic
representation in which wavy lines represent Green functions of 
$\pi_i^T$ and solid lines represent Green functions of $\phi$.
A line with one arrow is a retarded Green function, and a line with two
outgoing arrows represents a symmetric correlation function. The 
three-mode vertices correspond to the non-linear interaction terms in 
Eq.~(\ref{modH_1},\ref{modH_2}). Finally, the box-insertion in the 
correlation functions represents the strength of the noise in 
Eq.~(\ref{noise-phi},\ref{noise-pi}).

 The renormalization of $\eta$ was referred to as the ``stickiness of 
sound" in Ref.~\cite{Kovtun:2011np,Chafin:2012eq}. In the present case, we 
are not including sound modes and the stickiness of the fluid arises
only from the self-coupling of shear modes. Repeating the calculation 
\cite{Kovtun:2011np,Chafin:2012eq} for shear modes only gives
\begin{align}
\label{eta-R}
\eta_R = \eta  +  \frac{7}{60\pi^2} \frac{\rho T\Lambda}{\eta}\, ,
\end{align}
where $\eta$ is the bare viscosity, $\eta_R$ is the renormalized
viscosity, and $\Lambda$ is a momentum-space cutoff. For a lattice 
regularized theory we expect $\Lambda\simeq \pi/a$, where $a$ is the 
lattice spacing. This result has a number of interesting consequences.
It implies that the renormalized viscosity cannot become arbitrarily small.
As the bare viscosity is reduced, loop corrections become more important
and eventually dominate over the bare viscosity. Of course, in this 
regime the one-loop approximation is no longer reliable. It is nevertheless
instructive to estimate the minimum viscosity predicted by the one-loop
calculation. We find
\begin{align}
 \left.\eta_R\right|_{\it min}= 
   \sqrt{\frac{7}{15\pi}}\, \sqrt{\frac{\rho T}{a}}\, , 
\label{eta-min}   
\end{align}   
which shows that the minimum kinematic viscosity scales inversely with 
the mass density of the fluid. The minimum viscosity is fluctuation-driven
and increases with $T$. In the units adopted here, $T=a=1$, we obtain 
$\eta_R|_{\it min}\simeq 0.39\sqrt{\rho}$.

 Viscosity is also renormalized by the coupling to the scalar field,
see Fig.~\ref{fig:diags}(b). This diagram involves $\phi\phi
\pi^T_i$ vertices from both Eq.~(\ref{modH_1}) and Eq.~(\ref{modH_2}). 
The coupling in Eq.~(\ref{modH_2}) is third order in gradients, and
goes beyond the usual Navier-Stokes approximation. As a consequence,
the renormalization of $\eta$ by the coupling to the $\phi$ field
has not been studied before. We find that the $(\nabla\phi)^2$ term 
in the free energy acts as an ultra-violet regulator, and the diagram
is UV finite. The result is sensitive to the bare correlation length
$\xi^0=m^{-1}$. We obtain
\begin{align}
  \eta_{R}=\eta + \frac{1}{160\pi} \frac{T\xi^0}{\Gamma}\, .
\label{eta-R-phi}  
\end{align}
For $\xi^0\sim a$ and $a=\Gamma=T=1$ we get $\Delta\eta_R\simeq 0.002$, 
which is a much smaller correction than Eq.~(\ref{eta-R}).

 The one-loop correction shown in Fig.~\ref{fig:diags}(a) not only 
renormalizes the viscosity, but also qualitatively changes the long 
time behavior of the correlation function. The diagram in
Fig.~\ref{fig:diags}(a) corresponds to the splitting of a diffusive
shear mode with wave number $k$ into two modes with wave number 
$k/2\pm p$. This gives a long time tail that scales as $t^{-3/2}
\exp(-D_\eta k^2t/2)$, where $D_\eta=\eta/\rho$. This tail is 
comparable to the tree level result in Eq.~(\ref{C-pi-lin}) for 
$t>3/(D_\eta k^2)\log(1/(t_0D_\eta k^2))$, where $t_0$ is a 
microscopic time scale. In this regime higher loop corrections 
are not suppressed, and the correlation function becomes 
non-perturbative. Delacretaz conjectured that $C_\pi\sim 
\exp(-\sqrt{\alpha D_\eta k^2 t})$ where $\alpha$ is a constant 
that has to be determined non-perturbatively \cite{Delacretaz:2020nit}.

\subsection{Dynamics: Order parameter}
\label{sec:dyn-crit}

 The correlation function of the order parameter is  
\begin{align}
 C_\phi(t,\vec{k}) = \left\langle \phi(0,\vec{k})
    \phi(t,-\vec{k}) \right\rangle \, .
\label{C_t}
\end{align}  
This correlation function can be used to define a wave-number dependent
relaxation rate $C(t,\vec{k})\sim \exp(-\Gamma_kt)$. A simple model for 
$\Gamma_k$ was proposed by Kawasaki \cite{Kawasaki:1970}
\begin{align}
\Gamma_k = \frac{\Gamma}{\xi^4} 
      \left(k\xi\right)^2 \left(1+(k\xi)^2\right)
     + \frac{T}{6\pi\eta_R\xi^3}\, K(k\xi)\, , 
\label{Gamk:Kaw}
\end{align}
where the Kawasaki function is given by \cite{Kawasaki:1970,Onuki:2002}
\begin{align}
   K(x) = \frac{3}{4} \left[ 
     1+x^2+\left(x^3-x^{-1}\right)\tan^{-1}(x)\right]\, , 
\label{Kaw}
\end{align}
which satisfies $K(x)\simeq x^2$ for $x\ll 1$ and $K(x)\simeq (3\pi/8)
x^3$ for $x\gg 1$. The first term in Eq.~(\ref{Gamk:Kaw}) is the 
mean field relaxation rate in a purely diffusive theory (model B). 
It predicts that the relaxation rate of a fluctuation with wave
number $k=k^*\sim \xi^{-1}$ scales as $\Gamma_{k^*}\sim \xi^{-4}$,
which corresponds to a dynamical critical exponent $z=4$. This 
value is close to the non-perturbative result in model B, $z=4-\eta$
(see Eq.~(\ref{def:eta})).

 The second term in Eq.~(\ref{Gamk:Kaw}) comes from the coupling
of the order parameter to the momentum density of the fluid, see
Fig.~\ref{fig:diags}(c) and (d). In the Kawasaki approximation 
this diagram is computed by assuming the damping rate of the 
shear mode to be of the form $(\eta_R/\rho)k^2$, and the damping 
rate of the order parameter to be of the mean field form, $\Gamma
k^2(\xi^{-2}+k^2)$. In this approximation the diagrams in 
Fig.~\ref{fig:diags}(c,d) are UV and IR finite, and the loop
integral scales as $\Gamma_k\sim T/(\eta_R\xi^3)$. Equation~(\ref{Gamk:Kaw})
then predicts that the dynamical exponent $z$ is equal to three. 
This is close to the result of a more systematic calculation in 
the $\epsilon$-expansion which, at two loops, predicts $z\simeq 
3.0712$ \cite{Adzhemyan:1999h}. However, compared to the 
$\epsilon$-expansion the Kawasaki approximation has the virtue of 
giving a simple prediction for the wave number dependence of the 
relaxation rate. At small $(k\xi)$ we have $\Gamma_k\sim k^2$, and
at large $(k\xi)$ we expect $\Gamma_k\sim k^3$.

 Once the full relaxation rate $\Gamma_k$ has been determined, one
can go back and check the assumption that shear modes are damped
as $(\eta_R/\rho)k^2$. This amounts to computing the diagrams
in Fig.~\ref{fig:diags}(b) with the Kawasaki expression for
$\Gamma_k$. The result is \cite{Hohenberg:1977ym}
\begin{align}
\label{eta-crit}
\eta_R = \eta \left[ 1 
     + \frac{8}{15\pi^2}\log(\xi/\xi_0)\right], 
\end{align}
which implies that the viscosity diverges at the critical point 
as $\eta_R\sim \xi^{x_\eta}$ with $x_\eta=8/(15\pi^2)\simeq 0.054$. 
This result is also close to that of the $\epsilon$-expansion, which
predicts (at two loops) $x_\eta\simeq 0.071$ \cite{Adzhemyan:1999h}.
The main observation is that $x_\eta$ is very small, so that the 
divergence in $\eta_R$ is weak, and the Kawasaki approximation is 
approximately self-consistent.

 Finally, we note that there is a more physical argument that supports
the prediction $z\simeq 3$. This argument can be found in the review 
of Hohenberg and Halperin \cite{Hohenberg:1977ym}, who attribute it 
to Arcovito et al.~\cite{Arcovito:1969}. First, we observe that in the 
regime $k\xi<1$ the relaxation rate in Eq.~(\ref{Gamk:Kaw}) scales as
$\Gamma_k\sim k^2$. This means that we can define a renormalized 
conductivity $\Gamma_R$ by the relation $\Gamma_k=\Gamma_R\chi^{-1}k^2$.
The critical scaling of $\Gamma_R$ defines a new scaling exponent 
$\Gamma_R\sim \xi^{x_\Gamma}$ \footnote{
Note that in the literature $x_\Gamma$ is usually denoted by 
$x_\lambda$.}.
The Kawasaki approximation predicts $x_\Gamma=1$.
  
 The second part of the argument is the observation that there is 
a physical picture that constrains $x_\eta+x_\Gamma$. Consider a 
non-zero external field $h(x)$. Gradients of $h(x)$ will drive a 
diffusive current $\vec\jmath = -\Gamma_R\vec\nabla h$. However, in 
a fluid we can also have convection. The mechanical force per unit 
volume is $\vec{f}_{\it mech}=-\phi\vec\nabla h$, and this force can 
act coherently in a volume $V\sim\xi^d$ controlled by the correlation 
length. The force $\vec{f}_{\it mech}$ is balanced by Stokes drag 
$\vec{f}_{\it visc}\sim \eta_R \xi^{d-2}\vec{v}$, where $\vec{v}$ 
is the drift velocity of the volume element. The forces balance if 
$\vec{v}\sim -(\phi \xi^2/\eta_R) \vec\nabla h$, and the resulting 
current is $\vec\jmath\sim - (\phi^2\xi^2/\eta_R)\vec\nabla h$.
Finally, the mean value of $\phi^2$ in the volume element is 
$\phi^2\sim \chi/\xi^d$, and the convective current corresponds to 
a singular contribution to the conductivity $\Gamma_R\sim (\chi/\eta_R)
\xi^{2-d}$. Using the scaling of $\Gamma_R,\eta_R$ and $\chi$
with the correlation length we obtain
\begin{align}
    x_\Gamma+x_\eta = 4-d-\eta^*\, . 
\label{x-Gam-eta}    
\end{align}
We note that this result is consistent with the Kawasaki approximation.
Furthermore, if $x_\eta$ and $\eta^*$ are small, then $x_\Gamma\simeq 
4-d-\eta^* \simeq 1$ (in $d=3$). 

  The final step in the argument is to note that $x_\Gamma$ can 
be related to the dynamical exponent $z$. For this purpose we 
match the behavior of the relaxation rate at small $(k\xi)$, 
$\Gamma_k\simeq \Gamma_r\chi^{-1}k^2$, to the result at large 
$(k\xi)$, which is $\Gamma_k\simeq k^z$. Consistency for $(k\xi)
\sim 1$ requires 
\begin{align}
   z=4-x_\Gamma-\eta\, .
\label{z-Gam}   
\end{align}
Combining Eq.~(\ref{x-Gam-eta}) and (\ref{z-Gam}) gives $z=d+x_\eta$,
and for $x_\eta\simeq 0$ we obtain $z\simeq d$.

\begin{figure}[t]
\begin{center}
\includegraphics[width=\columnwidth]{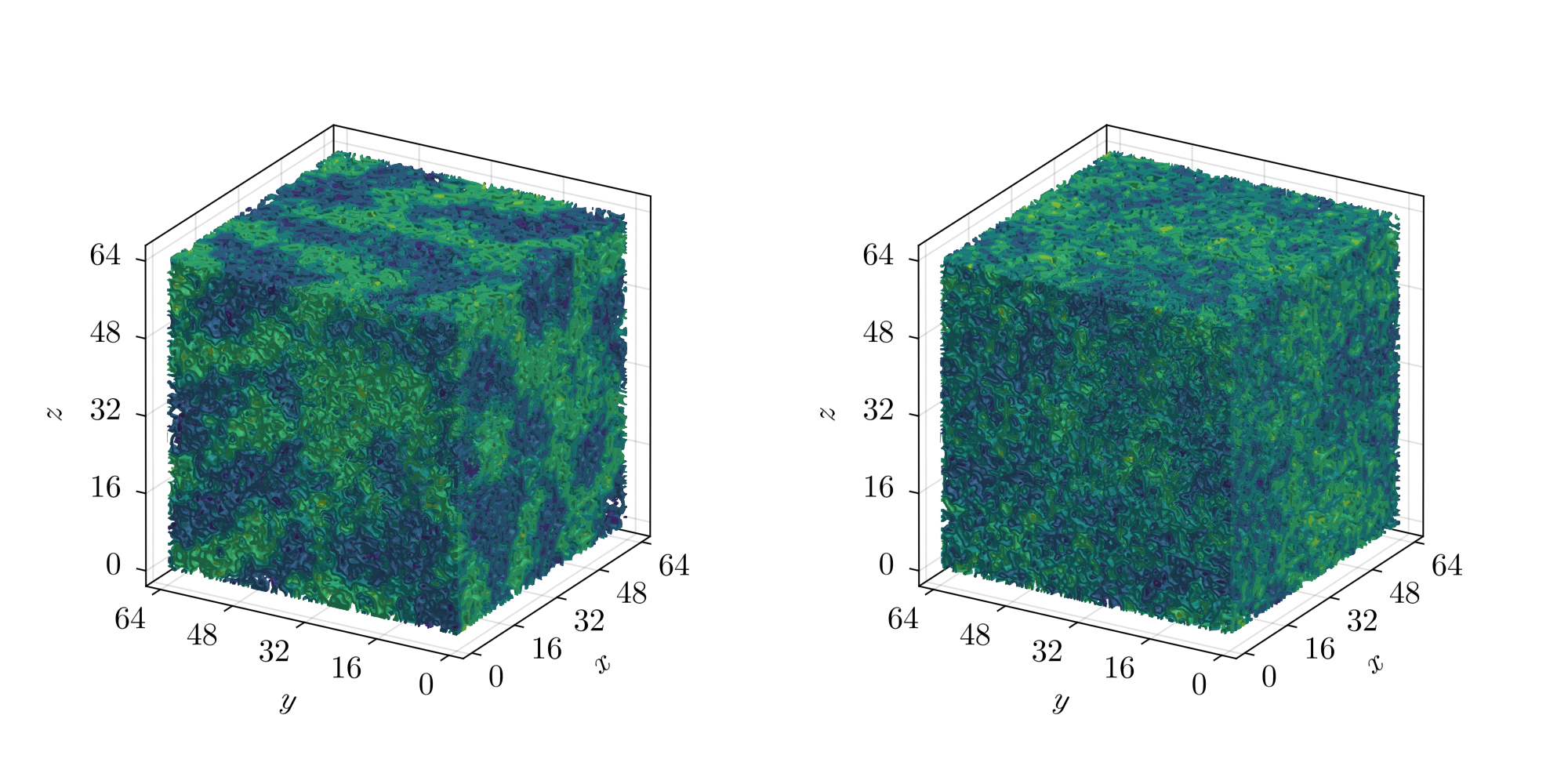}
\end{center}
\caption{Three-dimensional renderings of the order parameter 
$\phi(\vec{x},t)$ on a lattice $\vec{x}=\vec{n}a$ with volume 
$V=(La)^3$ where $L=64$. The left figure is obtained from a model 
H simulation below the critical temperature $m^2=m_c^2-1.1$, and
the right from a simulation near the critical point. Green/blue 
colors correspond to positive/negative values of $\phi$.}
\label{fig:3d}
\end{figure}

\section{Results}
\label{sec:res}

  We have simulated the algorithm described in Sect.~\ref{sec:num}
on cubic lattices of size $L^3$. The algorithm is essentially local,
and the computational cost for performing a single time step scales
as $L^3$. However, near a second order phase transition critical 
slowing down implies that of order $L^z$ time steps are required 
to equilibrate the system. In practice, near the critical point, 
we have been limited to lattices with $L^3\lesssim 64^3$ points.
Away from the critical point, or for simulations that are not close 
to equilibrium, much larger lattices are possible. Examples of
scalar field configurations encountered in model H simulations 
at and below the critical point are shown in Fig.~\ref{fig:3d}.

\subsection{Static properties}
\label{sec:statics}

\begin{figure}[t]
\begin{center}
\includegraphics[width=0.475\columnwidth]{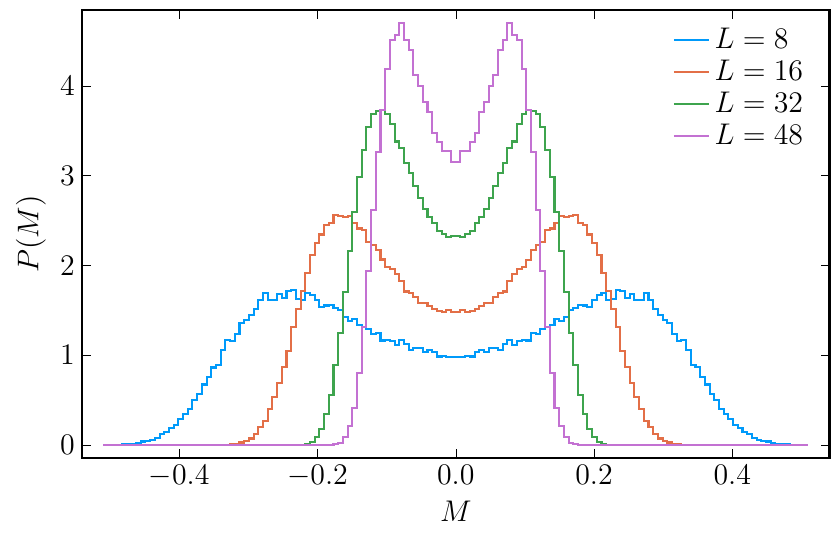}
\includegraphics[width=0.475\columnwidth]{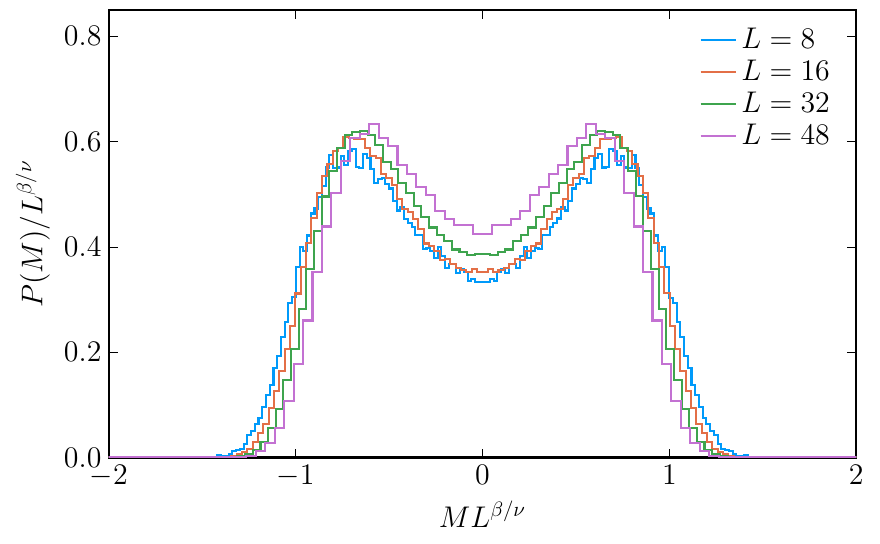}
\end{center}
\caption{The left panel shows the histogram of the magnetization 
$M$ at the critical point $m^2=m_c^2$ in model A for four different
volumes $L=8,16,32$ and 48. The figure in the right panel demonstrates
that the histogram scales with the exponent $\beta/\nu$.}
\label{fig:modA-hist}
\end{figure}

 The algorithm described in Section \ref{sec:diss} is designed 
to ensure that the equilibrium distribution $P[\phi,\pi^T_i]\sim
\exp(-{\cal H}[\phi,\pi^T_i]/T)$ is only governed by the Hamiltonian,
and does not depend on the transport coefficients or the Poisson
bracket terms. The Hamiltonian contains no coupling between 
the $\phi$ and $\pi_i^T$ fields, and as a result the equilibrium 
distribution of $\phi$ that emerges from the model H equations of 
motion (\ref{modH_1},\ref{modH_2}) is expected to be the same as 
the distribution obtained in model A \cite{Schaefer:2022bfm} and 
model B \cite{Chattopadhyay:2023jfm}. Here, model A is described 
by purely relaxational dynamics and model B is governed by diffusion,
but both models share the same Ising part of the Hamiltonian 
${\cal H}$. In a finite volume there is a difference between the 
model A distribution and the result in model B/H because the model A
dynamics does not conserve $\phi$. As a consequence, in model B/H
the integral of $\phi$ over the simulation volume cannot fluctuate. 

 Given that the static correlation functions of $\phi$ are the same 
in model A/B/H we can determine the location of the critical point 
using whichever dynamics is the most convenient. In practice the
main characteristic of the time evolution is the dynamical critical
exponent $z$. The dynamical exponent is smallest in model A, $z\simeq 
2$, which implies that this dynamical theory is the easiest to 
thermalize. In our previous work we determined the critical $m_c^2$ 
(for $\lambda=4$) by calculating Binder cumulants in model A 
\cite{Schaefer:2022bfm}. The cumulants are defined by 
\cite{Binder:1981}
\begin{align}
\label{Binder-def}
U  = 1 - \frac{\langle M^4 \rangle}{3\langle M^2 \rangle^2}\, ,  
\end{align}
where the magnetization $M$ is defined as 
\begin{align}
  M = \frac{1}{L^3}\sum_{\vec{x}} \phi(\vec{x})  \, . 
\end{align}
Binder cumulants are distinguished by the fact that at the infinite 
volume critical point $m_{c,\infty}^2\equiv m_c^2(L\to\infty)$ the 
leading finite volume corrections to $U$ cancel. We can then determine
$m_{c,\infty}^2$ by locating the point where the finite volume cumulants
cross. Furthermore, the critical value of $U$ is universal and has
been determined in high-precision simulations \cite{Hasenbusch:1998ve}.
Using this input we obtained $m_{c,\infty}^2 = -2.28587(7)$. A histogram
of the magnetization at the critical point in model A is shown in
Fig.~\ref{fig:modA-hist}. We observe that the distribution for 
different volume $V=L^3$ shows data collapse when the distribution 
is plotted as a function of $ML^{\beta/\nu}$, where $\beta\simeq 0.326$
is the order parameter exponent, and $\nu\simeq 0.630$ is the 
correlation length exponent.

\begin{figure}[t]
\begin{center}
\includegraphics[width=0.475\columnwidth]{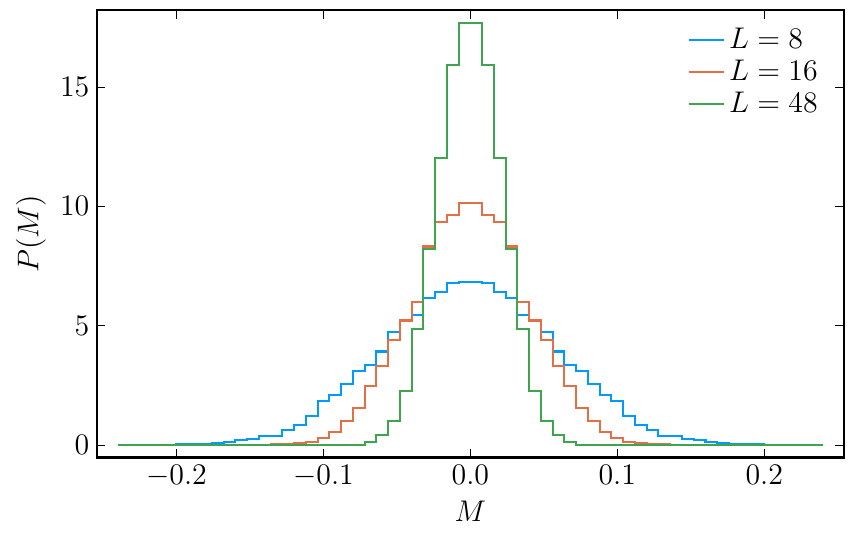}
\includegraphics[width=0.475\columnwidth]{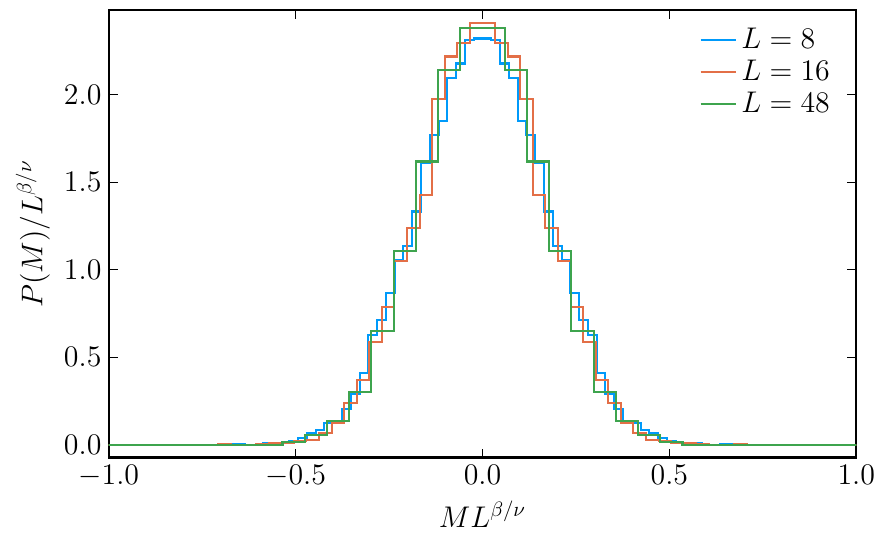}
\end{center}
\caption{The left panel shows the histogram of sub-volume magnetization
$M$ (integrated over half the simulation volume) at the critical point 
in model B for several different volumes. The distribution looks very
different from that for model A in Fig.~\ref{fig:modA-hist}, but the
scaling exponent is unchanged, as shown in the right panel. }
\label{fig:modB-hist}
\end{figure}

  In model B/H the total magnetization is conserved (typically we 
simulate at $M=0$), and the Binder cumulants are difficult to compute.
In principle one can study $M$ in some sub-volume, but then new finite
volume corrections due to charge conservation appear. In our prior
work on model B we studied the correlation length and relaxation time
as a function of $m^2-m_{c,\infty}^2$, and confirmed that the critical
point is the same in model A and B dynamics \cite{Chattopadhyay:2023jfm}.
This is supported by the results given in Fig.~\ref{fig:modB-hist}
which shows a histogram of the subvolume magnetization in model B 
at the critical point. We observe that the model B histograms 
differs significantly from the order parameter distribution in 
model A. We find, however, that scaling with $ML^{\beta/\nu}$ 
and the same critical exponents is satisfied. 

\begin{figure}[t]
\begin{center}
\includegraphics[width=0.6\columnwidth]{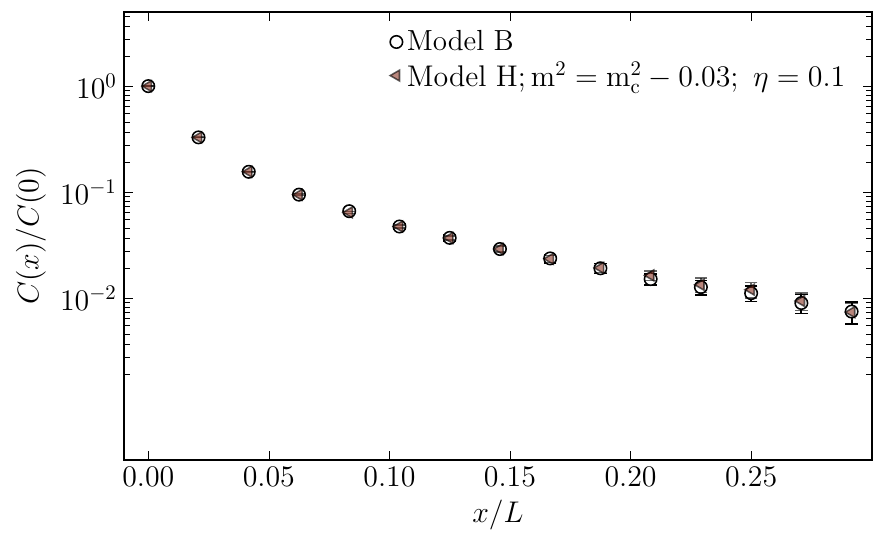}
\end{center}
\caption{Static correlation function of the order parameter 
$C(x)=\langle \phi(0,t)\phi(\vec{x},t)\rangle$ in model B 
and model H. We observe that a small shift in $m_c^2$ is required 
to reproduce the static model B correlation function in model H. 
}
\label{fig:static}
\end{figure}

In model H we have observed a small shift in the critical value of 
$m_c^2$. This is likely related to the fact that the model H advection
step described in Section \ref{sec:skew} does not exactly conserve 
the potential energy of $\phi$. As a consequence, the equilibrium value
of $\langle\phi^2\rangle$ is shifted by the advection step. In 
Fig.~\ref{fig:static} we show the static correlation function
\begin{align}
 C(x)=\langle \phi(0,t)\phi(\vec{x},t)\rangle   
\end{align}
in models B and H \footnote{
We compare model H to model B because the finite volume correction
to $C(x)$ due to charge conservation is the same in models B and H.}
We observe that a small shift, $\delta m_c^2=-0.03$ is required 
to reproduce the model B correlator using model H dynamics.

\subsection{Dynamic correlation function of the momentum density}
\label{sec:C-pi}

 We first consider the two-point function of the momentum density
defined in Eq.~(\ref{C-pi}). An example is shown in Fig.~\ref{fig:C-pi}
which shows the logarithmic derivative of $C_\pi$ in model H for two
different bare viscosities, $\eta=0.01$ and 0.05. We observe a 
well-defined plateau that can be used to extract the physical viscosity. 
We have performed similar calculations for a range of values of $\eta$ 
in a number of different theories. These include:

\begin{figure}[t]
\begin{center}
\includegraphics[width=0.70\columnwidth]{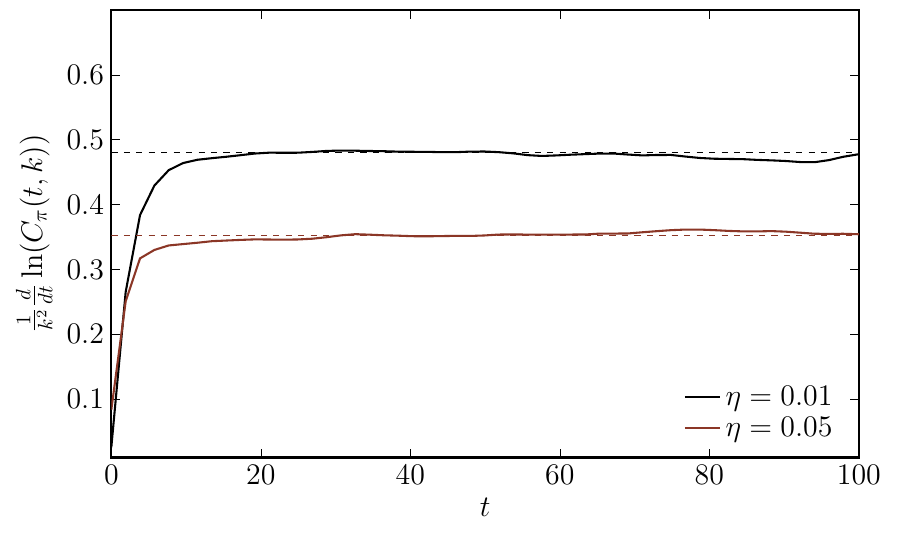}
\end{center}
\caption{Logarithmic derivative $k^{-2} d\log C_\pi(t)/dt$ of the 
two-point function of the momentum density $C_\pi(t,\vec{k})$ 
for the lowest non-zero
momentum mode on a lattice of size $L^3=48^3$.
The calculation was performed in model H for two different 
values of the bare viscosity, $\eta=0.01$ and 0.05. The dashed 
lines show the extracted value of the viscosity. }
\label{fig:C-pi}
\end{figure}

\begin{enumerate}
\item Model H as defined by Eqs.~(\ref{modH_1},\ref{modH_2}).
The discretized form of the advection terms is given in 
Eqs.~(\ref{phi_dot_discrete},\ref{pi_dot_discrete}), 
and the dissipative terms are defined in 
Eqs.~(\ref{phi-stoch-1}-\ref{pi-stoch-3}).
\item Model H0, which we have defined as Model H without the 
self-advection term in Eqs.~(\ref{modH_2}) and (\ref{pi_dot_discrete}).
As discussed in Sect.~\ref{sec:sound}, model H0 is a consistent
truncation of model H which is expected to be in the same dynamical
universality class.
\item Pure self advection, which takes into account the non-linear
dynamics of the momentum density, but ignores the coupling between 
$\phi$ and $\vec{\pi}^T$ in Eqs.~(\ref{modH_1},\ref{modH_2}) and
Eqs.~(\ref{phi_dot_discrete},\ref{pi_dot_discrete}). In this theory
the momentum density is decoupled from the critical dynamics of the 
order parameter. 
\item Pure momentum diffusion, which corresponds to ignoring all
mode couplings. In this approximation, the equation of the momentum
density is completely linear. 
\end{enumerate}

\begin{figure}[t]
\begin{center}
\includegraphics[width=0.70\columnwidth]{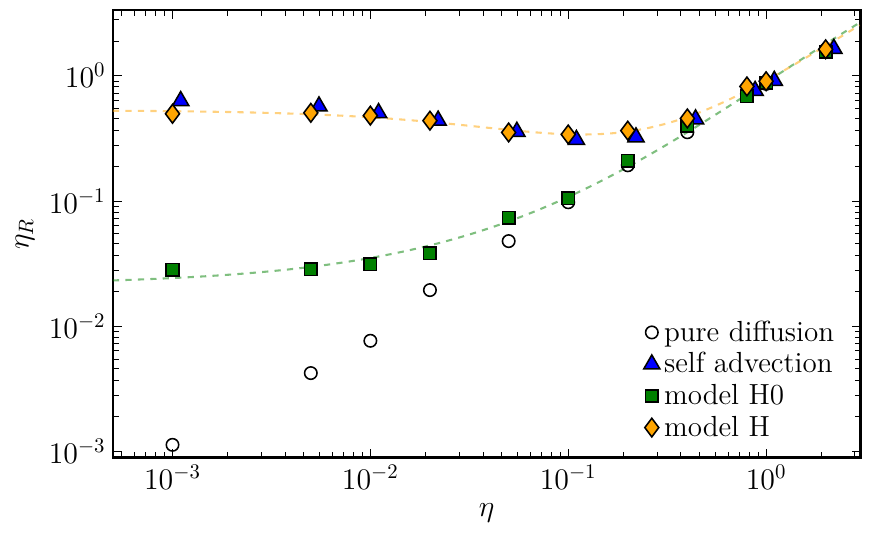}
\end{center}
\caption{
Physical viscosity as a function of the bare viscosity in four
different theories: 1) Model H at the critical point. 2) Model 
H0 at criticality (mutual advection of $\phi$ and $\vec\pi_T$ 
only). 3) Pure self-advection ($\pi^T$ only couples to itself).
4) Pure momentum diffusion (no mode-couplings). The results were 
obtained on a lattice of size $V=L^3$ with $L=48$, and the fluid
density was taken to be $\rho=1$. Self advection results are offset
horizontally to improve readability. The dashed lines show fits of 
the numerical data for model H0 and model H. 
}
\label{fig:visc}
\end{figure}

  The last of these theories, pure momentum diffusion, only serves
as a very basic check of our numerical procedure. Indeed, the open
circles in Fig.~\ref{fig:visc} show that the physical viscosity is 
very close to the bare viscosity. Pure self-advection takes into 
account the non-linear self coupling of the momentum density. In this 
case we observe a very significant renormalization of the viscosity.
The blue triangles in Fig.~\ref{fig:visc} show that the physical
viscosity tracks the bare viscosity down to $\eta\gtrsim 0.5$, but 
then it levels off and increases slightly for $\eta\lesssim
0.1$. This behavior is consistent with the one-loop renormalization
shown in Fig.~\ref{fig:diags}(a) and Eq.~(\ref{eta-R}). Indeed, 
even though the one-loop calculation is not reliable once the 
correction is comparable to the tree level term \footnote{
We have also not attempted to compute the diagram using a lattice
regulator, and have instead used the simple estimate $\Lambda
=\pi/a$ to relate the continuum cutoff $\lambda$ to the lattice 
spacing $a$.}
the observed viscosity minimum is quantitatively described by the
estimate $\eta_R|_{\it min}\simeq 0.39\sqrt{\rho}$ given in 
Sect.~\ref{sec:dyn-eta}.

 The green squares in Fig.~\ref{fig:visc} show that the renormalization
of the viscosity in model H0 is much smaller than the effect in model 
H or in a theory that includes only self-advection of the momentum
density. This is consistent with the expectations presented in 
Sect.~\ref{sec:dyn-eta}, where we argued that the renormalization 
of $\eta$ due to the coupling of $\vec{\pi}^T$ to itself, 
Eq.~(\ref{eta-R}), is much bigger than that due to the coupling 
to $\phi$, see Eq.~(\ref{eta-R-phi}). Note that the observed 
minimum in the viscosity, $\eta_R\simeq 3\cdot 10^{-2}$, 
is somewhat bigger than the prediction in Eq.~(\ref{eta-R-phi}). This
may be related to the fact that Fig.~\ref{fig:visc} shows the model H0
viscosity at the critical point, whereas Eq.~(\ref{eta-R-phi}) refers
to the non-critical background. Mode-coupling theories, as well as the 
$\epsilon$-expansion, predict a weak critical divergence in the model H0
viscosity, see Eq.~(\ref{eta-crit}). For the parameters considered 
here, critical effects are expected to lead to a multiplicative 
renormalization of order $\sim 10\%$. This effect may well be present 
for $\eta\lesssim 0.1$, but it is difficult to disentangle from the 
non-critical additive renormalization. Only a careful scaling analysis,
comparing different volumes as we will do in Sect.~\ref{sec:dyn-scal},
can be used to identify the critical scaling of $\eta_R$ \footnote{
Note that this will be difficult. In Sect.~\ref{sec:dyn-scal} we 
compare $L=40$ and $L=48$. Using $x_\lambda\simeq 0.05$ as predicted 
by the $\epsilon$-expansion, the expected enhancement of $\eta_R$ is 
$(48/40)^{0.05}\approx 1.01$, i.e. $1\%$ difference between the curves, 
too small to be observed with high confidence.}.
Finally, the yellow diamonds in Fig.~\ref{fig:visc} show the 
physical viscosity in model H \footnote{
We have performed fits of the renormalized viscosity as a function 
of the bare one using the trial function 
$ \ln \eta_R = f(x = \ln \eta) = d [ e\tanh(a x + b) - 1] + 
c (a x + b) [ 1 + \tanh(a x + b)]$. The fit yields $a =  0.28 (0.53)$, 
$b = 0.514 (0.81)$, $c =1.138 (0.87)$, $d = 1.86 (1.01)$ and $e = 1
(-0.372)$ for model H0 (model H).}. 
We observe that $\eta_R$ is dominated
by the non-critical renormalization that is already present in 
the calculation with pure self-advection. Compared to that, any 
critical enhancement is very difficult to observe. 

\subsection{Dynamic scaling}
\label{sec:dyn-scal}

  In this section we study the dynamical scaling behavior of the order
parameter correlation function $C(t,\vec{k})$ defined in Eq.~(\ref{C_t}).
Based on the discussion in Sect.~\ref{sec:dyn-crit} we expect that the
relaxation rate is modified by the coupling to the momentum density, 
and that this coupling changes the scaling of the relaxation rate from 
$\Gamma_k\sim \xi^{-4}$ to $\Gamma_k\sim\xi^{-3}$. The Kawasaki 
approximation given in Eq.~(\ref{Gamk:Kaw}) predicts that this crossover 
occurs for $\xi\gtrsim (6\pi\Gamma\eta_R)/T$, where $\eta_R$ is the 
physical viscosity shown in Fig.~\ref{fig:visc}. In model H we have 
$\eta_R|_{\it min}\simeq 0.3$ and this condition is only marginally 
satisfied. Testing dynamical scaling is therefore simpler in model
H0, where $\eta_R$ can be as small as $10^{-2}$.

\begin{figure}[t]
\begin{center}
\includegraphics[width=0.7\columnwidth]{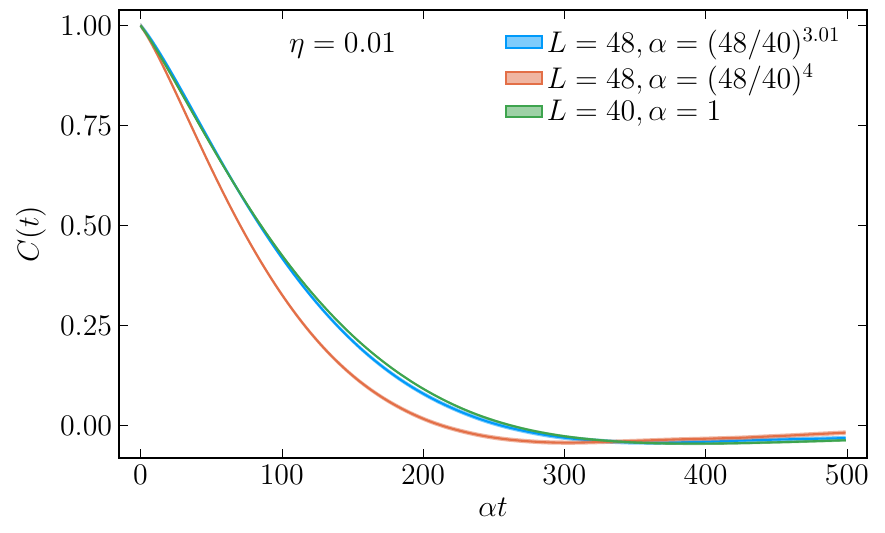}
\end{center}
\caption{
Normalized dynamic order parameter correlation function $C(t)
=C_\phi(t,k_n)/C_\phi(0,k_n)$ for the second non-trivial momentum 
mode $n=2$ and for different values of $L$ plotted as a function 
of the scaled time variable. This figure shows data taken at 
$\eta=10^{-2}$ for $L=40$ and $48$. Data collapse occurs for 
$z\simeq 3.01$, and the model B value $z\simeq 4$ is clearly 
excluded.
\label{fig:dyn-scal}}
\end{figure}

  Fig.~\ref{fig:dyn-scal} shows a calculation of the order parameter
correlation function in model H0 in two different volumes, $L=40$ and 
$L=48$, with a bare viscosity $\eta=10^{-2}$ and $\rho=1$. Dynamic 
scaling is the hypothesis that at the critical point the correlation 
function satisfies 
\begin{align}
    C_\phi(t,k;L) = \tilde{C}_\phi(t/L^z,kL) ,
\label{dyn:scal}    
\end{align}
where $\tilde{C}_\phi$ is a universal function, independent of $L$. 
Here, we have used the fact that at the critical point the correlation
length $\xi$ is only limited by the system size $L$. Away from the
critical point, when $a\ll\xi\ll L$, the dynamical scaling relation 
in Eq.~(\ref{dyn:scal}) holds with $L$ replaced by $\xi$. In
Fig.~\ref{fig:dyn-scal} we show $C_\phi(t,\vec{k})$ for the second 
moment mode on the lattice, $|\vec{k}|=4\pi/L$. This ensures that 
$|\vec{k}|L$ remains fixed as $L$ is varied. In the figure we look 
for dynamical scaling by plotting $C_\phi(\alpha t,kL)$ with 
$\alpha=1$ for $L=40$ and $\alpha=(40/48)^z$ for $L=48$. We observe
that dynamical scaling occurs for $z=3.01$. The data clearly exclude 
a scaling exponent close to four. 

\begin{figure}[t]
\begin{center}
\includegraphics[width=0.495\columnwidth]{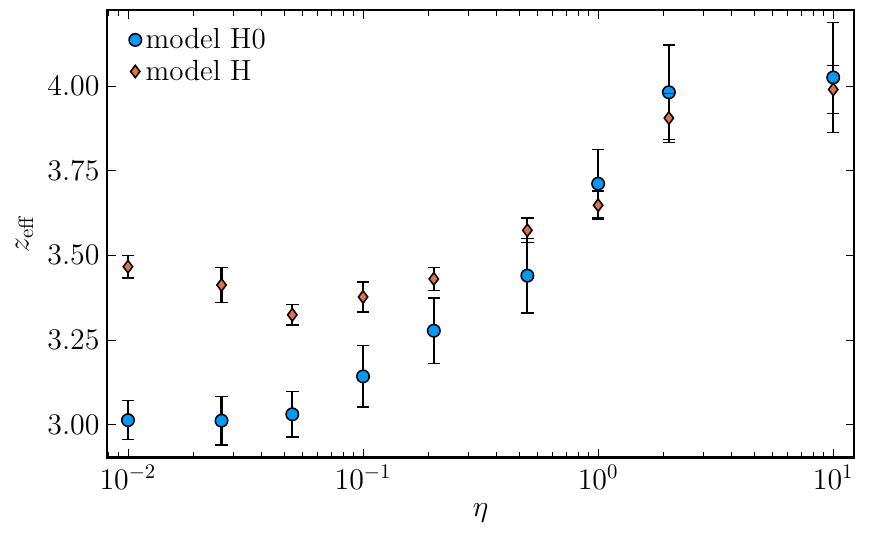}
\includegraphics[width=0.475\columnwidth]{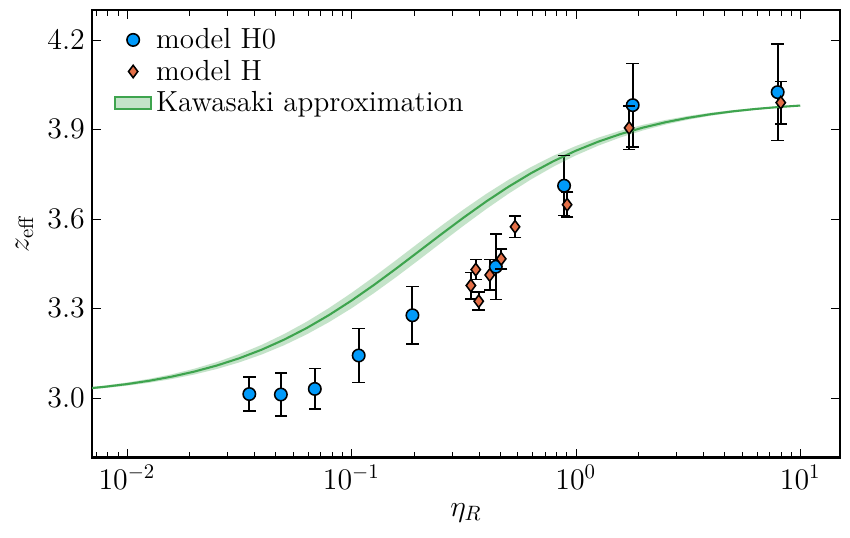}
\end{center}
\caption{Left panel: Dynamic scaling exponent $z_{\it eff}$ 
extracted from the correlation function $C(t,\vec{k})$ for different 
values of the bare and renormalized viscosity $\eta$ in model H0 
and H. We determined $z_{\it eff}$ by comparing the correlation 
function for two different volumes, $L=40$ and $48$. We also show 
the prediction of the Kawasaki approximation, Eq.~(\ref{Gamk:Kaw}). 
The error band is defined by varying the correlation length in the
range $\xi \in [L/2\pi, L/2]$. The right panel shows that $z_{\it 
eff}$ is only a function of the physical viscosity. 
\label{fig:z-eff}}
\end{figure}

 We have repeated this calculation for a range of values of $\eta$
in model H0. We extract an effective dynamical exponent $z_{\it eff}$ 
by minimizing the normalized correlation function $|C(t, L\!=\!40) -
C(\alpha t, L\!=\!48)|$ with respect to $\alpha=(40/48)^{z_{\it eff}}$ 
in the regime $C(t)>0.15$ (with $C(0)\equiv 1$). The results are shown
in Fig.~\ref{fig:z-eff}. We also show the results in a complete model 
H calculation. For small values of the bare viscosity the extracted 
$z_{\it eff}$ differs from the result in model H0. However, when
plotted against the renormalized viscosity extracted in 
Fig.~\ref{fig:visc} the dynamic exponent in the two theories agrees
(right panel in Fig.~\ref{fig:z-eff}). For comparison, we also show 
the prediction of the Kawasaki approximation $C(t,\vec{k})\sim\exp(-
\Gamma_k t)$ where the relaxation rate $\Gamma_k$ is given in 
Eq.~(\ref{Gamk:Kaw}). The value of $\xi$ at the critical point in 
a finite volume is not very well defined. In Fig.~\ref{fig:z-eff} 
we show the prediction of the Kawasaki approximation for $L/(2\pi)
<\xi<L/2$. 

   We observe that the effective dynamical exponent does indeed 
exhibit a crossover from $z_{\it eff}\sim 4$ to $z_{\it eff}\sim 3$,
and that this crossover is semi-quantitatively described by the 
Kawasaki approximation. For $\eta_R= 10^{-2}$ we find $z_{\it eff}=
3.013 \pm 0.058$, which is our best estimate of the dynamical 
exponent in the infinite volume limit. This value is consistent 
with the prediction of the $\epsilon$-expansion at two loops, 
$z \simeq 3.0712$ \cite{Adzhemyan:1999h}.
   
\subsection{Two-dimensional fluids}
\label{sec:2d}

 In this section we study the behavior of a two-dimensional
fluid near a critical point in the Ising universality class. 
There are several observations that motivate our study.  The
lower critical dimension of model $H$ is $d=2$, and interesting
non-perturbative phenomena may take place in two-dimensional
fluids. Also, the main source of theoretical information regarding 
the dynamical scaling exponents is the expansion around $d=4-
\epsilon$ dimensions, which is most questionable near $d=2$.
Calculations in $d=2$ may well be a good laboratory to compare
numerical simulation, the $\epsilon$-expansion, and the functional
renormalization group. Finally, two-dimensional fluids may have 
some relevance to the real world. For example, critical fluctuations
in $d=2$ may control the critical dynamics of a fluid with rapid
longitudinal expansion. 

\begin{figure}
\centering
\includegraphics[width=0.475\linewidth]{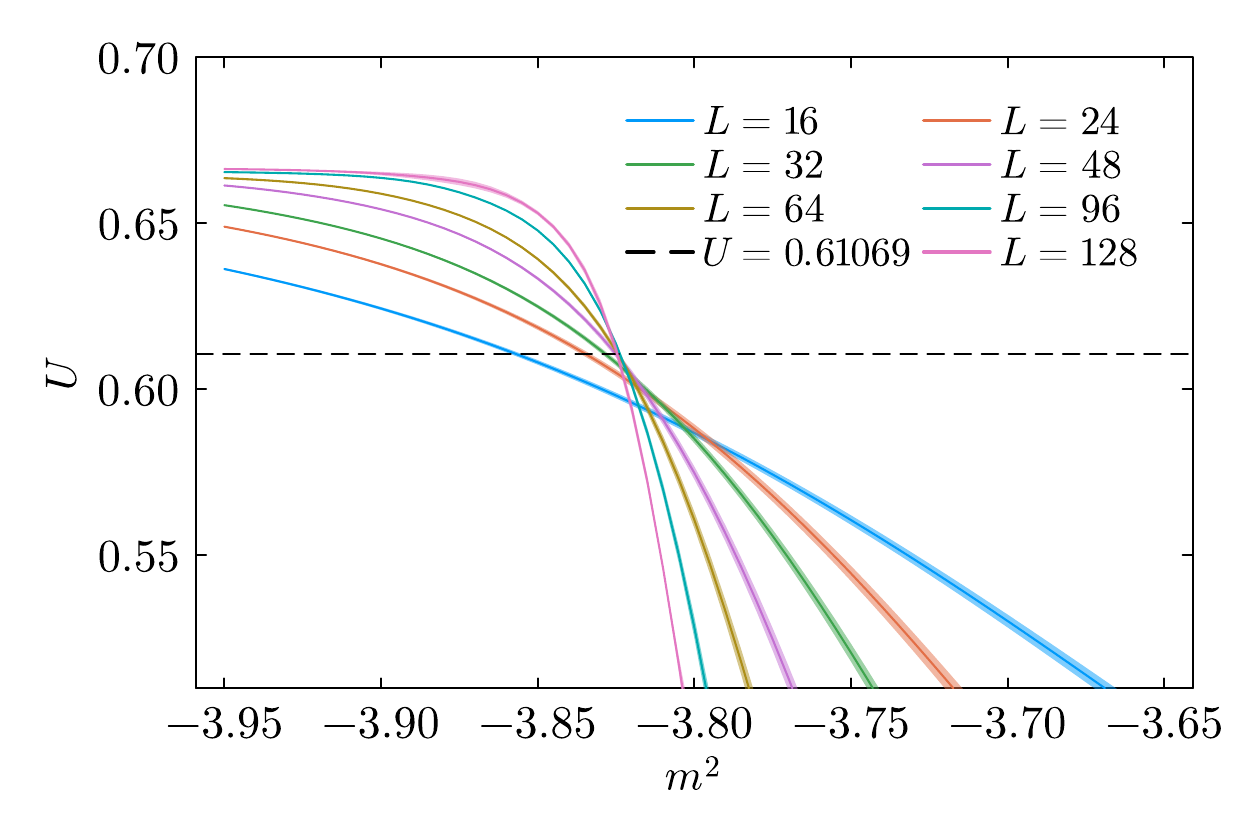}
\includegraphics[width=0.475\linewidth]{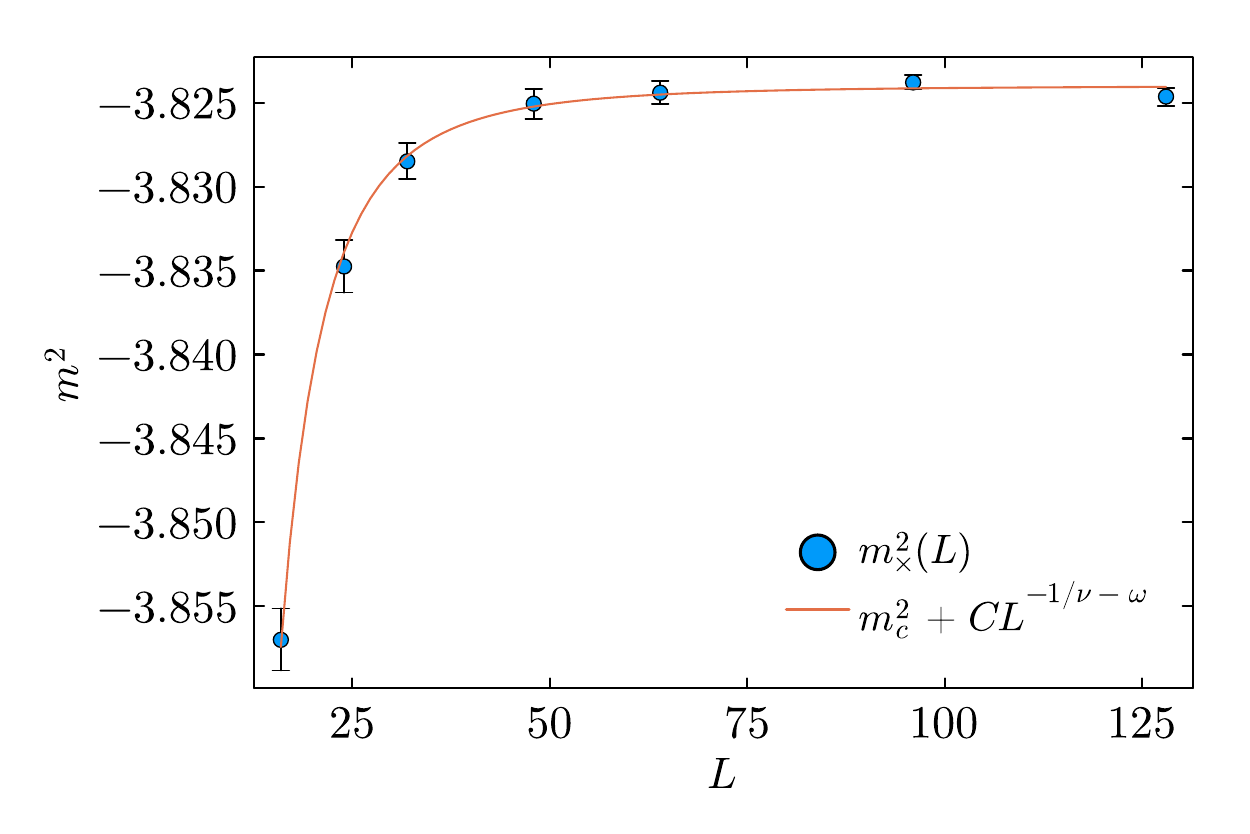}
\caption{Left panel: Binder cumulant for model A in two dimensions
as a function of the control parameter $m^2$ for different values 
of the system size $L$. The dashed line shows the critical value 
$U^*$ of the Binder cumulant in the infinite volume limit
\cite{Kamieniarz:1993}.
Right panel: The figure shows the location $m^2$ of the crossing
points $U(m^2,L)=U^*$ for different values of $L$, together 
with a fit of the form $m^2=m_c^2+CL^{1/\nu-\omega}$. Here,
$\omega=2$ is a finite size scaling exponent. The fit yields  
$m_c^2 = -3.8240 \pm 0.0003$. }
\label{fig:2d-Binder}
\end{figure}

 The static behavior of a two-dimensional fluid in the Ising 
universality class is well understood, both from theory and 
simulation. The free energy was determined by Onsager, and the
correlation functions are governed by a well known two-dimensional
conformal field theory \cite{Francesco:2012}. The correlation function
exponent is given by $\eta^*=0.25$, and the correlation length exponent 
is $\nu=1$. Less is known about the dynamics. The epsilon expansion 
predicts $z=2.179$ and $x_\eta=0.179$ \cite{Adzhemyan:1999h}. 
Calculations based on the functional renormalization group find 
similar values of $z$, but either zero \cite{Roth:2024hcu} or 
significantly smaller values \cite{Chen:2024lzz} for $x_\eta$.
A vanishing critical exponent for the shear viscosity can also be
deduced from mode coupling approximation \cite{Ohta:1975}. Note 
that there is a subtlety in trying to extract the viscosity of 
two-dimensional fluid. Even in a non-critical fluid there is a 
logarithmic divergence of the viscosity in the limit that the 
frequency goes to zero \cite{Kadanoff:1989,Kovtun:2012rj,
Delacretaz:2020nit}.
 
 The algorithm described in Sect.~\ref{sec:num} does not require 
any modifications in the two-dimensional case. In order to study 
critical dynamics we first have to locate the critical $m_c^2$. 
As explained in Sect.~\ref{sec:statics} this is most easily done
using model A dynamics. The left panel of Fig.~\ref{fig:2d-Binder}
shows the Binder cumulant $U$ defined in Eq.~(\ref{Binder-def}) for 
different values of the system size $L$ and the mass parameter 
$m^2$. The dashed line shows the known critical value of $U$ in 
two dimensions, $U^*=0.61069$ \cite{Kamieniarz:1993}. The right
panel shows the extrapolation to infinite $L$, which gives  
$m_c^2 =  -3.8240 \pm 0.0003$. Following the procedure described
in Sect.~\ref{sec:statics} we have checked whether there is a 
small shift in $m_c^2$ as we go from model A/B dynamics to model H.
We find that the model H critical point is at $m_c^2\simeq -3.859$.

\begin{figure}[t]
\begin{center}
\includegraphics[width=0.495\columnwidth]{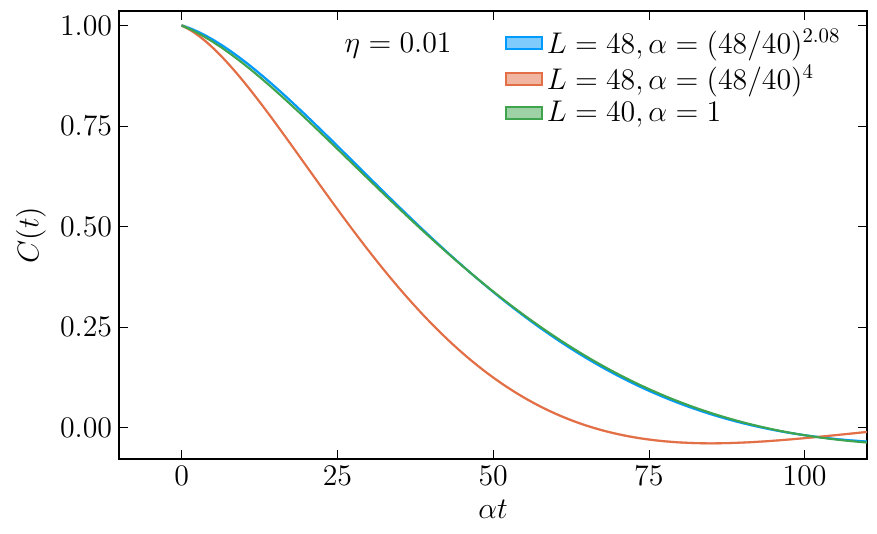}
\includegraphics[width=0.475\columnwidth]{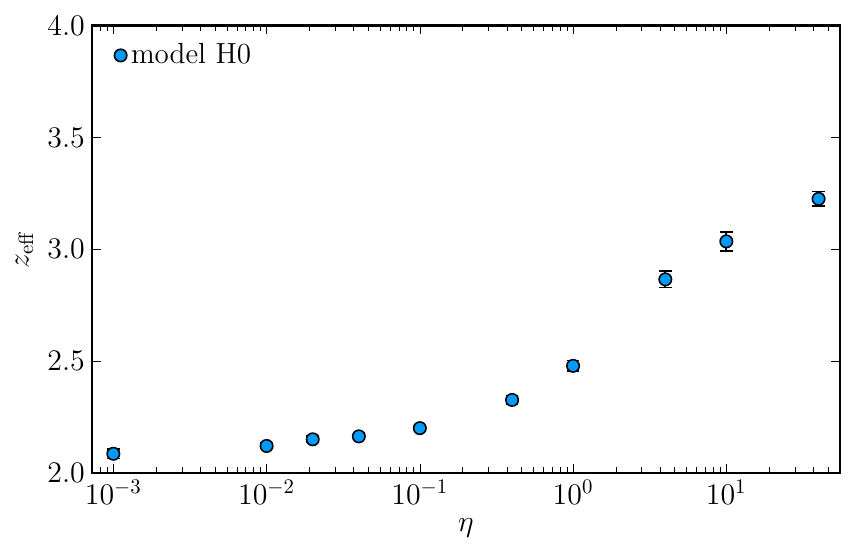}
\end{center}
\caption{Left panel: Dynamic scaling of the normalized order parameter 
correlation function $C(t)$for the second non-trivial momentum mode 
in model H0 in two dimensions, see Fig.~\ref{fig:dyn-scal} for the
corresponding result in three dimensions. Right panel: Dynamic 
scaling exponent $z_{\it eff}$ extracted from the correlation 
function $C(t,\vec{k})$ for different values of the bare viscosity 
$\eta$ in model H0  in $d=2$. 
\label{fig:z2d}}
\end{figure}

  Given the difficulty in extracting the shear viscosity of a 
two-dimensional fluid we have not attempted to study the 
renormalized viscosity in $2d$. In order to determine the dynamic 
scaling exponent we have computed the order parameter correlation
function in Eq.~(\ref{C_t}) for a range of values of the bare 
viscosity. As in Sect.~\ref{sec:dyn-scal} we look for dynamic 
scaling of the correlation functions computed for different values 
of $L$. In the left panel of Fig.~\ref{fig:z2d} we show data 
collapse for the dynamic correlation function in two volumes with
$L=40$ and $L=48$. For $\eta_0=10^{-2}$ we obtain $z=2.11\pm0.015$, 
quite consistent with the prediction of the $\epsilon$-expansion 
quoted above, $z=2.179$. 
  
  In the right panel of Fig.~\ref{fig:z2d} we show the dependence 
of the value of $z$ extracted in a finite volume on the bare 
viscosity. Similar to the results in three dimensions we observe 
a crossover, and the asymptotic value of $z\simeq 2$ is only 
reached for very small values of $\eta$. Contrary to the three
dimensional case we do not observe the model B value $z\simeq 3.75$ 
for the largest values of $\eta$ that we have explored. This is 
related to the fact that in two dimensions the value of $\eta$ 
needed for the shear modes to decouple scales as $\eta\sim \xi^2
T/\Gamma$, compared to $\xi T/\Gamma$ in three dimensions. Our
algorithm becomes inefficient for very large values of $\eta$, 
and we have not studied values of $\eta$ greater than $10^2$. We 
have verified that mean-field model B asymtotics is seen in smaller
volumes. 

\section{Summary and outlook}
\label{sec:sum}

 We have studied a new approach to stochastic fluid dynamics
which is based on a very simple and robust Metropolis algorithm. 
The main advantage of the algorithm is that fluctuation-dissipation
relations are automatically satisfied, so that the simulation 
converges to an equilibrium distribution which is sampled from 
the desired microscopic free energy functional. We have used this 
method to compute the dynamic critical exponents of two and 
three-dimensional fluids near a critical endpoint in the universality
class of the Ising model. 

 There are a number of issues that we would like to address in the 
future. One class of problems involves improved calculations of 
the correlation function investigated in the present work. For 
example, what is the true long-time behavior of the momentum 
density correlation function? How can we best extract the macroscopic
shear viscosity of the fluid? What is the best way to measure 
the scaling behavior of transport coefficients near the critical
point? Addressing these issues may benefit from semi-analytical 
studies, for example based on Dyson-Schwinger equations or the 
functional renormalization group. 

 The second class of challenges has to do with making the 
simulations more realistic. This involves going to compressible
fluid dynamics and including a realistic equation of state.
In a compressible simulation we can study the behavior of the 
speed of sound, the sound attenuation rate, and the bulk 
viscosity. This is straightforward in principle, but we will 
have to study the renormalization of the equation of state. 

 Third, we would like to study non-trivial background flows,
for example systems that undergo both longitudinal and 
transverse expansion. We are particularly interested in 
genuinely relativistic flows, in which the expansion velocity 
reaches a significant fraction of the speed of light. In this
context it has recently been argued that the density frame 
formulation of hydrodynamics \cite{Armas:2020mpr,Basar:2024qxd,
Bhambure:2024gnf} is particularly well suited for stochastic 
fluid dynamics and the Metropolis algorithm. 
 
{\it Acknowledgments:}
This work is supported by the U.S. Department of Energy, Office of
Science, Office of Nuclear Physics through the Contracts 
DE-FG02-03ER41260, DE-SC0024622 and DE-SC0020081.  This work
used computing resources provided by the NC State University 
High Performance Computing Services Core Facility (RRID:SCR-022168), 
as well as resources funded by the  Wesley O.~Doggett endowment. 
We thank Andrew Petersen for assistance in working with the HPC
infrastructure. 

\noindent
Codes used to generate the data are available at 
\cite{Ott:2025a,Ott:2025b}.

\bibliographystyle{apsrev4-1}
\bibliography{bib}

\begin{thebibliography}{79}%
\makeatletter
\providecommand \@ifxundefined [1]{%
 \@ifx{#1\undefined}
}%
\providecommand \@ifnum [1]{%
 \ifnum #1\expandafter \@firstoftwo
 \else \expandafter \@secondoftwo
 \fi
}%
\providecommand \@ifx [1]{%
 \ifx #1\expandafter \@firstoftwo
 \else \expandafter \@secondoftwo
 \fi
}%
\providecommand \natexlab [1]{#1}%
\providecommand \enquote  [1]{``#1''}%
\providecommand \bibnamefont  [1]{#1}%
\providecommand \bibfnamefont [1]{#1}%
\providecommand \citenamefont [1]{#1}%
\providecommand \href@noop [0]{\@secondoftwo}%
\providecommand \href [0]{\begingroup \@sanitize@url \@href}%
\providecommand \@href[1]{\@@startlink{#1}\@@href}%
\providecommand \@@href[1]{\endgroup#1\@@endlink}%
\providecommand \@sanitize@url [0]{\catcode `\\12\catcode `\$12\catcode
  `\&12\catcode `\#12\catcode `\^12\catcode `\_12\catcode `\%12\relax}%
\providecommand \@@startlink[1]{}%
\providecommand \@@endlink[0]{}%
\providecommand \url  [0]{\begingroup\@sanitize@url \@url }%
\providecommand \@url [1]{\endgroup\@href {#1}{\urlprefix }}%
\providecommand \urlprefix  [0]{URL }%
\providecommand \Eprint [0]{\href }%
\providecommand \doibase [0]{http://dx.doi.org/}%
\providecommand \selectlanguage [0]{\@gobble}%
\providecommand \bibinfo  [0]{\@secondoftwo}%
\providecommand \bibfield  [0]{\@secondoftwo}%
\providecommand \translation [1]{[#1]}%
\providecommand \BibitemOpen [0]{}%
\providecommand \bibitemStop [0]{}%
\providecommand \bibitemNoStop [0]{.\EOS\space}%
\providecommand \EOS [0]{\spacefactor3000\relax}%
\providecommand \BibitemShut  [1]{\csname bibitem#1\endcsname}%
\let\auto@bib@innerbib\@empty
\bibitem [{\citenamefont {Sch\"afer}\ and\ \citenamefont
  {Teaney}(2009)}]{Schafer:2009dj}%
  \BibitemOpen
  \bibfield  {author} {\bibinfo {author} {\bibfnamefont {T.}~\bibnamefont
  {Sch\"afer}}\ and\ \bibinfo {author} {\bibfnamefont {D.}~\bibnamefont
  {Teaney}},\ }\href {\doibase 10.1088/0034-4885/72/12/126001} {\bibfield
  {journal} {\bibinfo  {journal} {Rept. Prog. Phys.}\ }\textbf {\bibinfo
  {volume} {72}},\ \bibinfo {pages} {126001} (\bibinfo {year} {2009})},\
  \Eprint {http://arxiv.org/abs/0904.3107} {arXiv:0904.3107 [hep-ph]}
  \BibitemShut {NoStop}%
\bibitem [{\citenamefont {Jeon}\ and\ \citenamefont
  {Heinz}(2015)}]{Jeon:2015dfa}%
  \BibitemOpen
  \bibfield  {author} {\bibinfo {author} {\bibfnamefont {S.}~\bibnamefont
  {Jeon}}\ and\ \bibinfo {author} {\bibfnamefont {U.}~\bibnamefont {Heinz}},\
  }\href {\doibase 10.1142/S0218301315300106} {\bibfield  {journal} {\bibinfo
  {journal} {Int. J. Mod. Phys. E}\ }\textbf {\bibinfo {volume} {24}},\
  \bibinfo {pages} {1530010} (\bibinfo {year} {2015})},\ \Eprint
  {http://arxiv.org/abs/1503.03931} {arXiv:1503.03931 [hep-ph]} \BibitemShut
  {NoStop}%
\bibitem [{\citenamefont {Romatschke}\ and\ \citenamefont
  {Romatschke}(2019)}]{Romatschke:2017ejr}%
  \BibitemOpen
  \bibfield  {author} {\bibinfo {author} {\bibfnamefont {P.}~\bibnamefont
  {Romatschke}}\ and\ \bibinfo {author} {\bibfnamefont {U.}~\bibnamefont
  {Romatschke}},\ }\href {\doibase 10.1017/9781108651998} {\emph {\bibinfo
  {title} {{Relativistic Fluid Dynamics In and Out of Equilibrium}}}},\
  Cambridge Monographs on Mathematical Physics\ (\bibinfo  {publisher}
  {Cambridge University Press},\ \bibinfo {year} {2019})\ \Eprint
  {http://arxiv.org/abs/1712.05815} {arXiv:1712.05815 [nucl-th]} \BibitemShut
  {NoStop}%
\bibitem [{\citenamefont {Landau}\ and\ \citenamefont
  {Lifshitz}(1980)}]{landau:SMII}%
  \BibitemOpen
  \bibfield  {author} {\bibinfo {author} {\bibfnamefont {L.}~\bibnamefont
  {Landau}}\ and\ \bibinfo {author} {\bibfnamefont {E.}~\bibnamefont
  {Lifshitz}},\ }\href@noop {} {\emph {\bibinfo {title} {Course of Theoretical
  Physics: Statistical Physics,Part 2}}}\ (\bibinfo  {publisher} {Pergamon
  Press},\ \bibinfo {year} {1980})\BibitemShut {NoStop}%
\bibitem [{\citenamefont {Kovtun}\ \emph {et~al.}(2011)\citenamefont {Kovtun},
  \citenamefont {Moore},\ and\ \citenamefont {Romatschke}}]{Kovtun:2011np}%
  \BibitemOpen
  \bibfield  {author} {\bibinfo {author} {\bibfnamefont {P.}~\bibnamefont
  {Kovtun}}, \bibinfo {author} {\bibfnamefont {G.~D.}\ \bibnamefont {Moore}}, \
  and\ \bibinfo {author} {\bibfnamefont {P.}~\bibnamefont {Romatschke}},\
  }\href@noop {} {\bibfield  {journal} {\bibinfo  {journal} {Phys. Rev. D}\
  }\textbf {\bibinfo {volume} {84}},\ \bibinfo {pages} {025006} (\bibinfo
  {year} {2011})},\ \Eprint {http://arxiv.org/abs/1104.1586} {arXiv:1104.1586
  [hep-ph]} \BibitemShut {NoStop}%
\bibitem [{\citenamefont {Crossley}\ \emph {et~al.}(2017)\citenamefont
  {Crossley}, \citenamefont {Glorioso},\ and\ \citenamefont
  {Liu}}]{Crossley:2015evo}%
  \BibitemOpen
  \bibfield  {author} {\bibinfo {author} {\bibfnamefont {M.}~\bibnamefont
  {Crossley}}, \bibinfo {author} {\bibfnamefont {P.}~\bibnamefont {Glorioso}},
  \ and\ \bibinfo {author} {\bibfnamefont {H.}~\bibnamefont {Liu}},\ }\href
  {\doibase 10.1007/JHEP09(2017)095} {\bibfield  {journal} {\bibinfo  {journal}
  {JHEP}\ }\textbf {\bibinfo {volume} {09}},\ \bibinfo {pages} {095} (\bibinfo
  {year} {2017})},\ \Eprint {http://arxiv.org/abs/1511.03646} {arXiv:1511.03646
  [hep-th]} \BibitemShut {NoStop}%
\bibitem [{\citenamefont {Chen-Lin}\ \emph {et~al.}(2019)\citenamefont
  {Chen-Lin}, \citenamefont {Delacr\'etaz},\ and\ \citenamefont
  {Hartnoll}}]{Chen-Lin:2018kfl}%
  \BibitemOpen
  \bibfield  {author} {\bibinfo {author} {\bibfnamefont {X.}~\bibnamefont
  {Chen-Lin}}, \bibinfo {author} {\bibfnamefont {L.~V.}\ \bibnamefont
  {Delacr\'etaz}}, \ and\ \bibinfo {author} {\bibfnamefont {S.~A.}\
  \bibnamefont {Hartnoll}},\ }\href {\doibase 10.1103/PhysRevLett.122.091602}
  {\bibfield  {journal} {\bibinfo  {journal} {Phys. Rev. Lett.}\ }\textbf
  {\bibinfo {volume} {122}},\ \bibinfo {pages} {091602} (\bibinfo {year}
  {2019})},\ \Eprint {http://arxiv.org/abs/1811.12540} {arXiv:1811.12540
  [hep-th]} \BibitemShut {NoStop}%
\bibitem [{\citenamefont {Jain}\ and\ \citenamefont
  {Kovtun}(2022)}]{Jain:2020zhu}%
  \BibitemOpen
  \bibfield  {author} {\bibinfo {author} {\bibfnamefont {A.}~\bibnamefont
  {Jain}}\ and\ \bibinfo {author} {\bibfnamefont {P.}~\bibnamefont {Kovtun}},\
  }\href {\doibase 10.1103/PhysRevLett.128.071601} {\bibfield  {journal}
  {\bibinfo  {journal} {Phys. Rev. Lett.}\ }\textbf {\bibinfo {volume} {128}},\
  \bibinfo {pages} {071601} (\bibinfo {year} {2022})},\ \Eprint
  {http://arxiv.org/abs/2009.01356} {arXiv:2009.01356 [hep-th]} \BibitemShut
  {NoStop}%
\bibitem [{\citenamefont {Delacretaz}(2020)}]{Delacretaz:2020nit}%
  \BibitemOpen
  \bibfield  {author} {\bibinfo {author} {\bibfnamefont {L.~V.}\ \bibnamefont
  {Delacretaz}},\ }\href {\doibase 10.21468/SciPostPhys.9.3.034} {\bibfield
  {journal} {\bibinfo  {journal} {SciPost Phys.}\ }\textbf {\bibinfo {volume}
  {9}},\ \bibinfo {pages} {034} (\bibinfo {year} {2020})},\ \Eprint
  {http://arxiv.org/abs/2006.01139} {arXiv:2006.01139 [hep-th]} \BibitemShut
  {NoStop}%
\bibitem [{\citenamefont {Chao}\ and\ \citenamefont
  {Sch{\"a}fer}(2021)}]{Chao:2020kcf}%
  \BibitemOpen
  \bibfield  {author} {\bibinfo {author} {\bibfnamefont {J.}~\bibnamefont
  {Chao}}\ and\ \bibinfo {author} {\bibfnamefont {T.}~\bibnamefont
  {Sch{\"a}fer}},\ }\href {\doibase 10.1007/JHEP01(2021)071} {\bibfield
  {journal} {\bibinfo  {journal} {JHEP}\ }\textbf {\bibinfo {volume} {01}},\
  \bibinfo {pages} {071} (\bibinfo {year} {2021})},\ \Eprint
  {http://arxiv.org/abs/2008.01269} {arXiv:2008.01269 [hep-th]} \BibitemShut
  {NoStop}%
\bibitem [{\citenamefont {Basar}(2024)}]{Basar:2024srd}%
  \BibitemOpen
  \bibfield  {author} {\bibinfo {author} {\bibfnamefont {G.}~\bibnamefont
  {Basar}},\ }\href@noop {} {\  (\bibinfo {year} {2024})},\ \Eprint
  {http://arxiv.org/abs/2410.02866} {arXiv:2410.02866 [hep-th]} \BibitemShut
  {NoStop}%
\bibitem [{\citenamefont {Hohenberg}\ and\ \citenamefont
  {Halperin}(1977)}]{Hohenberg:1977ym}%
  \BibitemOpen
  \bibfield  {author} {\bibinfo {author} {\bibfnamefont {P.~C.}\ \bibnamefont
  {Hohenberg}}\ and\ \bibinfo {author} {\bibfnamefont {B.~I.}\ \bibnamefont
  {Halperin}},\ }\href {\doibase 10.1103/RevModPhys.49.435} {\bibfield
  {journal} {\bibinfo  {journal} {Rev. Mod. Phys.}\ }\textbf {\bibinfo {volume}
  {49}},\ \bibinfo {pages} {435} (\bibinfo {year} {1977})}\BibitemShut
  {NoStop}%
\bibitem [{\citenamefont {Rajagopal}\ and\ \citenamefont
  {Wilczek}(1993)}]{Rajagopal:1992qz}%
  \BibitemOpen
  \bibfield  {author} {\bibinfo {author} {\bibfnamefont {K.}~\bibnamefont
  {Rajagopal}}\ and\ \bibinfo {author} {\bibfnamefont {F.}~\bibnamefont
  {Wilczek}},\ }\href {\doibase 10.1016/0550-3213(93)90502-G} {\bibfield
  {journal} {\bibinfo  {journal} {Nucl. Phys. B}\ }\textbf {\bibinfo {volume}
  {399}},\ \bibinfo {pages} {395} (\bibinfo {year} {1993})},\ \Eprint
  {http://arxiv.org/abs/hep-ph/9210253} {arXiv:hep-ph/9210253} \BibitemShut
  {NoStop}%
\bibitem [{\citenamefont {Son}\ and\ \citenamefont
  {Stephanov}(2004)}]{Son:2004iv}%
  \BibitemOpen
  \bibfield  {author} {\bibinfo {author} {\bibfnamefont {D.~T.}\ \bibnamefont
  {Son}}\ and\ \bibinfo {author} {\bibfnamefont {M.~A.}\ \bibnamefont
  {Stephanov}},\ }\href@noop {} {\bibfield  {journal} {\bibinfo  {journal}
  {Phys. Rev. D}\ }\textbf {\bibinfo {volume} {70}},\ \bibinfo {pages} {056001}
  (\bibinfo {year} {2004})},\ \Eprint {http://arxiv.org/abs/hep-ph/0401052}
  {arXiv:hep-ph/0401052} \BibitemShut {NoStop}%
\bibitem [{\citenamefont {Arcovito}\ \emph {et~al.}(1969)\citenamefont
  {Arcovito}, \citenamefont {Faloci}, \citenamefont {Roberti},\ and\
  \citenamefont {Mistura}}]{Arcovito:1969}%
  \BibitemOpen
  \bibfield  {author} {\bibinfo {author} {\bibfnamefont {G.}~\bibnamefont
  {Arcovito}}, \bibinfo {author} {\bibfnamefont {C.}~\bibnamefont {Faloci}},
  \bibinfo {author} {\bibfnamefont {M.}~\bibnamefont {Roberti}}, \ and\
  \bibinfo {author} {\bibfnamefont {L.}~\bibnamefont {Mistura}},\ }\href
  {\doibase 10.1103/PhysRevLett.22.1040} {\bibfield  {journal} {\bibinfo
  {journal} {Phys. Rev. Lett.}\ }\textbf {\bibinfo {volume} {22}},\ \bibinfo
  {pages} {1040} (\bibinfo {year} {1969})}\BibitemShut {NoStop}%
\bibitem [{\citenamefont {Canet}\ and\ \citenamefont
  {Chate}(2007)}]{Canet:2006xu}%
  \BibitemOpen
  \bibfield  {author} {\bibinfo {author} {\bibfnamefont {L.}~\bibnamefont
  {Canet}}\ and\ \bibinfo {author} {\bibfnamefont {H.}~\bibnamefont {Chate}},\
  }\href {\doibase 10.1088/1751-8113/40/9/002} {\bibfield  {journal} {\bibinfo
  {journal} {J. Phys.}\ }\textbf {\bibinfo {volume} {40}},\ \bibinfo {pages}
  {1937} (\bibinfo {year} {2007})},\ \Eprint
  {http://arxiv.org/abs/cond-mat/0610468} {arXiv:cond-mat/0610468} \BibitemShut
  {NoStop}%
\bibitem [{\citenamefont {Canet}\ \emph {et~al.}(2011)\citenamefont {Canet},
  \citenamefont {Chate},\ and\ \citenamefont {Delamotte}}]{Canet:2011wf}%
  \BibitemOpen
  \bibfield  {author} {\bibinfo {author} {\bibfnamefont {L.}~\bibnamefont
  {Canet}}, \bibinfo {author} {\bibfnamefont {H.}~\bibnamefont {Chate}}, \ and\
  \bibinfo {author} {\bibfnamefont {B.}~\bibnamefont {Delamotte}},\ }\href
  {\doibase 10.1088/1751-8113/44/49/495001} {\bibfield  {journal} {\bibinfo
  {journal} {J. Phys. A}\ }\textbf {\bibinfo {volume} {44}},\ \bibinfo {pages}
  {495001} (\bibinfo {year} {2011})},\ \Eprint {http://arxiv.org/abs/1106.4129}
  {arXiv:1106.4129 [cond-mat.stat-mech]} \BibitemShut {NoStop}%
\bibitem [{\citenamefont {Mesterh{\'a}zy}\ \emph {et~al.}(2013)\citenamefont
  {Mesterh{\'a}zy}, \citenamefont {Stockemer}, \citenamefont {Palhares},\ and\
  \citenamefont {Berges}}]{Mesterhazy:2013naa}%
  \BibitemOpen
  \bibfield  {author} {\bibinfo {author} {\bibfnamefont {D.}~\bibnamefont
  {Mesterh{\'a}zy}}, \bibinfo {author} {\bibfnamefont {J.~H.}\ \bibnamefont
  {Stockemer}}, \bibinfo {author} {\bibfnamefont {L.~F.}\ \bibnamefont
  {Palhares}}, \ and\ \bibinfo {author} {\bibfnamefont {J.}~\bibnamefont
  {Berges}},\ }\href {\doibase 10.1103/PhysRevB.88.174301} {\bibfield
  {journal} {\bibinfo  {journal} {Phys. Rev. B}\ }\textbf {\bibinfo {volume}
  {88}},\ \bibinfo {pages} {174301} (\bibinfo {year} {2013})},\ \Eprint
  {http://arxiv.org/abs/1307.1700} {arXiv:1307.1700 [cond-mat.stat-mech]}
  \BibitemShut {NoStop}%
\bibitem [{\citenamefont {Chen}\ \emph
  {et~al.}(2024{\natexlab{a}})\citenamefont {Chen}, \citenamefont {Tan},\ and\
  \citenamefont {Fu}}]{Chen:2023tqc}%
  \BibitemOpen
  \bibfield  {author} {\bibinfo {author} {\bibfnamefont {Y.-r.}\ \bibnamefont
  {Chen}}, \bibinfo {author} {\bibfnamefont {Y.-y.}\ \bibnamefont {Tan}}, \
  and\ \bibinfo {author} {\bibfnamefont {W.-j.}\ \bibnamefont {Fu}},\ }\href
  {\doibase 10.1103/PhysRevD.109.094044} {\bibfield  {journal} {\bibinfo
  {journal} {Phys. Rev. D}\ }\textbf {\bibinfo {volume} {109}},\ \bibinfo
  {pages} {094044} (\bibinfo {year} {2024}{\natexlab{a}})},\ \Eprint
  {http://arxiv.org/abs/2312.05870} {arXiv:2312.05870 [hep-ph]} \BibitemShut
  {NoStop}%
\bibitem [{\citenamefont {Roth}\ and\ \citenamefont {von
  Smekal}(2023)}]{Roth:2023wbp}%
  \BibitemOpen
  \bibfield  {author} {\bibinfo {author} {\bibfnamefont {J.~V.}\ \bibnamefont
  {Roth}}\ and\ \bibinfo {author} {\bibfnamefont {L.}~\bibnamefont {von
  Smekal}},\ }\href {\doibase 10.1007/JHEP10(2023)065} {\bibfield  {journal}
  {\bibinfo  {journal} {JHEP}\ }\textbf {\bibinfo {volume} {10}},\ \bibinfo
  {pages} {065} (\bibinfo {year} {2023})},\ \Eprint
  {http://arxiv.org/abs/2303.11817} {arXiv:2303.11817 [hep-ph]} \BibitemShut
  {NoStop}%
\bibitem [{\citenamefont {Chen}\ \emph
  {et~al.}(2024{\natexlab{b}})\citenamefont {Chen}, \citenamefont {Tan},\ and\
  \citenamefont {Fu}}]{Chen:2024lzz}%
  \BibitemOpen
  \bibfield  {author} {\bibinfo {author} {\bibfnamefont {Y.-R.}\ \bibnamefont
  {Chen}}, \bibinfo {author} {\bibfnamefont {Y.-Y.}\ \bibnamefont {Tan}}, \
  and\ \bibinfo {author} {\bibfnamefont {W.-J.}\ \bibnamefont {Fu}},\
  }\href@noop {} {\  (\bibinfo {year} {2024}{\natexlab{b}})},\ \Eprint
  {http://arxiv.org/abs/2406.00679} {arXiv:2406.00679 [hep-ph]} \BibitemShut
  {NoStop}%
\bibitem [{\citenamefont {Roth}\ \emph {et~al.}(2024)\citenamefont {Roth},
  \citenamefont {Ye}, \citenamefont {Schlichting},\ and\ \citenamefont {von
  Smekal}}]{Roth:2024hcu}%
  \BibitemOpen
  \bibfield  {author} {\bibinfo {author} {\bibfnamefont {J.~V.}\ \bibnamefont
  {Roth}}, \bibinfo {author} {\bibfnamefont {Y.}~\bibnamefont {Ye}}, \bibinfo
  {author} {\bibfnamefont {S.}~\bibnamefont {Schlichting}}, \ and\ \bibinfo
  {author} {\bibfnamefont {L.}~\bibnamefont {von Smekal}},\ }\href@noop {} {\
  (\bibinfo {year} {2024})},\ \Eprint {http://arxiv.org/abs/2409.14470}
  {arXiv:2409.14470 [hep-ph]} \BibitemShut {NoStop}%
\bibitem [{\citenamefont {Kapusta}\ \emph {et~al.}(2012)\citenamefont
  {Kapusta}, \citenamefont {Muller},\ and\ \citenamefont
  {Stephanov}}]{Kapusta:2011gt}%
  \BibitemOpen
  \bibfield  {author} {\bibinfo {author} {\bibfnamefont {J.~I.}\ \bibnamefont
  {Kapusta}}, \bibinfo {author} {\bibfnamefont {B.}~\bibnamefont {Muller}}, \
  and\ \bibinfo {author} {\bibfnamefont {M.}~\bibnamefont {Stephanov}},\ }\href
  {\doibase 10.1103/PhysRevC.85.054906} {\bibfield  {journal} {\bibinfo
  {journal} {Phys. Rev. C}\ }\textbf {\bibinfo {volume} {85}},\ \bibinfo
  {pages} {054906} (\bibinfo {year} {2012})},\ \Eprint
  {http://arxiv.org/abs/1112.6405} {arXiv:1112.6405 [nucl-th]} \BibitemShut
  {NoStop}%
\bibitem [{\citenamefont {Mukherjee}\ \emph {et~al.}(2016)\citenamefont
  {Mukherjee}, \citenamefont {Venugopalan},\ and\ \citenamefont
  {Yin}}]{Mukherjee:2016kyu}%
  \BibitemOpen
  \bibfield  {author} {\bibinfo {author} {\bibfnamefont {S.}~\bibnamefont
  {Mukherjee}}, \bibinfo {author} {\bibfnamefont {R.}~\bibnamefont
  {Venugopalan}}, \ and\ \bibinfo {author} {\bibfnamefont {Y.}~\bibnamefont
  {Yin}},\ }\href {\doibase 10.1103/PhysRevLett.117.222301} {\bibfield
  {journal} {\bibinfo  {journal} {Phys. Rev. Lett.}\ }\textbf {\bibinfo
  {volume} {117}},\ \bibinfo {pages} {222301} (\bibinfo {year} {2016})},\
  \Eprint {http://arxiv.org/abs/1605.09341} {arXiv:1605.09341 [hep-ph]}
  \BibitemShut {NoStop}%
\bibitem [{\citenamefont {Akamatsu}\ \emph {et~al.}(2017)\citenamefont
  {Akamatsu}, \citenamefont {Mazeliauskas},\ and\ \citenamefont
  {Teaney}}]{Akamatsu:2016llw}%
  \BibitemOpen
  \bibfield  {author} {\bibinfo {author} {\bibfnamefont {Y.}~\bibnamefont
  {Akamatsu}}, \bibinfo {author} {\bibfnamefont {A.}~\bibnamefont
  {Mazeliauskas}}, \ and\ \bibinfo {author} {\bibfnamefont {D.}~\bibnamefont
  {Teaney}},\ }\href@noop {} {\bibfield  {journal} {\bibinfo  {journal} {Phys.
  Rev. C}\ }\textbf {\bibinfo {volume} {95}},\ \bibinfo {pages} {014909}
  (\bibinfo {year} {2017})},\ \Eprint {http://arxiv.org/abs/1606.07742}
  {arXiv:1606.07742 [nucl-th]} \BibitemShut {NoStop}%
\bibitem [{\citenamefont {Stephanov}\ and\ \citenamefont
  {Yin}(2018)}]{Stephanov:2017ghc}%
  \BibitemOpen
  \bibfield  {author} {\bibinfo {author} {\bibfnamefont {M.}~\bibnamefont
  {Stephanov}}\ and\ \bibinfo {author} {\bibfnamefont {Y.}~\bibnamefont
  {Yin}},\ }\href@noop {} {\bibfield  {journal} {\bibinfo  {journal} {Phys.
  Rev. D}\ }\textbf {\bibinfo {volume} {98}},\ \bibinfo {pages} {036006}
  (\bibinfo {year} {2018})},\ \Eprint {http://arxiv.org/abs/1712.10305}
  {arXiv:1712.10305 [nucl-th]} \BibitemShut {NoStop}%
\bibitem [{\citenamefont {Akamatsu}\ \emph {et~al.}(2019)\citenamefont
  {Akamatsu}, \citenamefont {Teaney}, \citenamefont {Yan},\ and\ \citenamefont
  {Yin}}]{Akamatsu:2018vjr}%
  \BibitemOpen
  \bibfield  {author} {\bibinfo {author} {\bibfnamefont {Y.}~\bibnamefont
  {Akamatsu}}, \bibinfo {author} {\bibfnamefont {D.}~\bibnamefont {Teaney}},
  \bibinfo {author} {\bibfnamefont {F.}~\bibnamefont {Yan}}, \ and\ \bibinfo
  {author} {\bibfnamefont {Y.}~\bibnamefont {Yin}},\ }\href {\doibase
  10.1103/PhysRevC.100.044901} {\bibfield  {journal} {\bibinfo  {journal}
  {Phys. Rev. C}\ }\textbf {\bibinfo {volume} {100}},\ \bibinfo {pages}
  {044901} (\bibinfo {year} {2019})},\ \Eprint
  {http://arxiv.org/abs/1811.05081} {arXiv:1811.05081 [nucl-th]} \BibitemShut
  {NoStop}%
\bibitem [{\citenamefont {Martinez}\ and\ \citenamefont
  {Sch\"afer}(2019)}]{Martinez:2018wia}%
  \BibitemOpen
  \bibfield  {author} {\bibinfo {author} {\bibfnamefont {M.}~\bibnamefont
  {Martinez}}\ and\ \bibinfo {author} {\bibfnamefont {T.}~\bibnamefont
  {Sch\"afer}},\ }\href@noop {} {\bibfield  {journal} {\bibinfo  {journal}
  {Phys. Rev. C}\ }\textbf {\bibinfo {volume} {99}},\ \bibinfo {pages} {054902}
  (\bibinfo {year} {2019})},\ \Eprint {http://arxiv.org/abs/1812.05279}
  {arXiv:1812.05279 [hep-th]} \BibitemShut {NoStop}%
\bibitem [{\citenamefont {An}\ \emph {et~al.}(2019)\citenamefont {An},
  \citenamefont {Ba\c{s}ar}, \citenamefont {Stephanov},\ and\ \citenamefont
  {Yee}}]{An:2019osr}%
  \BibitemOpen
  \bibfield  {author} {\bibinfo {author} {\bibfnamefont {X.}~\bibnamefont
  {An}}, \bibinfo {author} {\bibfnamefont {G.}~\bibnamefont {Ba\c{s}ar}},
  \bibinfo {author} {\bibfnamefont {M.}~\bibnamefont {Stephanov}}, \ and\
  \bibinfo {author} {\bibfnamefont {H.-U.}\ \bibnamefont {Yee}},\ }\href@noop
  {} {\bibfield  {journal} {\bibinfo  {journal} {Phys. Rev. C}\ }\textbf
  {\bibinfo {volume} {100}},\ \bibinfo {pages} {024910} (\bibinfo {year}
  {2019})},\ \Eprint {http://arxiv.org/abs/1902.09517} {arXiv:1902.09517
  [hep-th]} \BibitemShut {NoStop}%
\bibitem [{\citenamefont {An}\ \emph {et~al.}(2020)\citenamefont {An},
  \citenamefont {Ba\c{s}ar}, \citenamefont {Stephanov},\ and\ \citenamefont
  {Yee}}]{An:2019csj}%
  \BibitemOpen
  \bibfield  {author} {\bibinfo {author} {\bibfnamefont {X.}~\bibnamefont
  {An}}, \bibinfo {author} {\bibfnamefont {G.}~\bibnamefont {Ba\c{s}ar}},
  \bibinfo {author} {\bibfnamefont {M.}~\bibnamefont {Stephanov}}, \ and\
  \bibinfo {author} {\bibfnamefont {H.-U.}\ \bibnamefont {Yee}},\ }\href@noop
  {} {\bibfield  {journal} {\bibinfo  {journal} {Phys. Rev. C}\ }\textbf
  {\bibinfo {volume} {102}},\ \bibinfo {pages} {034901} (\bibinfo {year}
  {2020})},\ \Eprint {http://arxiv.org/abs/1912.13456} {arXiv:1912.13456
  [hep-th]} \BibitemShut {NoStop}%
\bibitem [{\citenamefont {An}\ \emph {et~al.}(2021)\citenamefont {An},
  \citenamefont {Ba\c{s}ar}, \citenamefont {Stephanov},\ and\ \citenamefont
  {Yee}}]{An:2020vri}%
  \BibitemOpen
  \bibfield  {author} {\bibinfo {author} {\bibfnamefont {X.}~\bibnamefont
  {An}}, \bibinfo {author} {\bibfnamefont {G.}~\bibnamefont {Ba\c{s}ar}},
  \bibinfo {author} {\bibfnamefont {M.}~\bibnamefont {Stephanov}}, \ and\
  \bibinfo {author} {\bibfnamefont {H.-U.}\ \bibnamefont {Yee}},\ }\href
  {\doibase 10.1103/PhysRevLett.127.072301} {\bibfield  {journal} {\bibinfo
  {journal} {Phys. Rev. Lett.}\ }\textbf {\bibinfo {volume} {127}},\ \bibinfo
  {pages} {072301} (\bibinfo {year} {2021})},\ \Eprint
  {http://arxiv.org/abs/2009.10742} {arXiv:2009.10742 [hep-th]} \BibitemShut
  {NoStop}%
\bibitem [{\citenamefont {Bell}\ \emph {et~al.}(2007)\citenamefont {Bell},
  \citenamefont {Garcia},\ and\ \citenamefont {Williams}}]{Bell:2007}%
  \BibitemOpen
  \bibfield  {author} {\bibinfo {author} {\bibfnamefont {J.~B.}\ \bibnamefont
  {Bell}}, \bibinfo {author} {\bibfnamefont {A.~L.}\ \bibnamefont {Garcia}}, \
  and\ \bibinfo {author} {\bibfnamefont {S.~A.}\ \bibnamefont {Williams}},\
  }\href {\doibase 10.1103/PhysRevE.76.016708} {\bibfield  {journal} {\bibinfo
  {journal} {Phys. Rev. E}\ }\textbf {\bibinfo {volume} {76}},\ \bibinfo
  {pages} {016708} (\bibinfo {year} {2007})},\ \Eprint
  {http://arxiv.org/abs/arXiv:math/0612324 [math.NA]} {arXiv:math/0612324
  [math.NA]} \BibitemShut {NoStop}%
\bibitem [{\citenamefont {Donev}\ \emph {et~al.}(2010)\citenamefont {Donev},
  \citenamefont {Vanden-Eijnden}, \citenamefont {Garcia},\ and\ \citenamefont
  {Bell}}]{Donev:2010}%
  \BibitemOpen
  \bibfield  {author} {\bibinfo {author} {\bibfnamefont {A.}~\bibnamefont
  {Donev}}, \bibinfo {author} {\bibfnamefont {E.}~\bibnamefont
  {Vanden-Eijnden}}, \bibinfo {author} {\bibfnamefont {A.}~\bibnamefont
  {Garcia}}, \ and\ \bibinfo {author} {\bibfnamefont {J.}~\bibnamefont
  {Bell}},\ }\href {\doibase 10.2140/camcos.2010.5.149} {\bibfield  {journal}
  {\bibinfo  {journal} {Communications in Applied Mathematics and Computational
  Science}\ }\textbf {\bibinfo {volume} {5}},\ \bibinfo {pages} {149} (\bibinfo
  {year} {2010})},\ \Eprint {http://arxiv.org/abs/arXiv:0906.2425
  [physics.flu-dyn]} {arXiv:0906.2425 [physics.flu-dyn]} \BibitemShut {NoStop}%
\bibitem [{\citenamefont {Camley}\ and\ \citenamefont
  {Brown}(2010)}]{Camley:2010}%
  \BibitemOpen
  \bibfield  {author} {\bibinfo {author} {\bibfnamefont {B.~A.}\ \bibnamefont
  {Camley}}\ and\ \bibinfo {author} {\bibfnamefont {F.~L.~H.}\ \bibnamefont
  {Brown}},\ }\href {\doibase 10.1103/PhysRevLett.105.148102} {\bibfield
  {journal} {\bibinfo  {journal} {Phys. Rev. Lett.}\ }\textbf {\bibinfo
  {volume} {105}},\ \bibinfo {pages} {148102} (\bibinfo {year} {2010})},\
  \Eprint {http://arxiv.org/abs/arXiv:1105.4898 [cond-mat.soft]}
  {arXiv:1105.4898 [cond-mat.soft]} \BibitemShut {NoStop}%
\bibitem [{\citenamefont {Balboa}\ \emph {et~al.}(2012)\citenamefont {Balboa},
  \citenamefont {Bell}, \citenamefont {Delgado-Buscalioni}, \citenamefont
  {Donev}, \citenamefont {Fai}, \citenamefont {Griffith},\ and\ \citenamefont
  {Peskin}}]{Balboa:2012}%
  \BibitemOpen
  \bibfield  {author} {\bibinfo {author} {\bibfnamefont {F.}~\bibnamefont
  {Balboa}}, \bibinfo {author} {\bibfnamefont {J.~B.}\ \bibnamefont {Bell}},
  \bibinfo {author} {\bibfnamefont {R.}~\bibnamefont {Delgado-Buscalioni}},
  \bibinfo {author} {\bibfnamefont {A.}~\bibnamefont {Donev}}, \bibinfo
  {author} {\bibfnamefont {T.~G.}\ \bibnamefont {Fai}}, \bibinfo {author}
  {\bibfnamefont {B.~E.}\ \bibnamefont {Griffith}}, \ and\ \bibinfo {author}
  {\bibfnamefont {C.~S.}\ \bibnamefont {Peskin}},\ }\href@noop {} {\bibfield
  {journal} {\bibinfo  {journal} {Multiscale Modeling \& Simulation (SIAM)}\
  }\textbf {\bibinfo {volume} {10}},\ \bibinfo {pages} {1369} (\bibinfo {year}
  {2012})},\ \Eprint {http://arxiv.org/abs/arXiv:1108.5188 [physics.flu-dyn]}
  {arXiv:1108.5188 [physics.flu-dyn]} \BibitemShut {NoStop}%
\bibitem [{\citenamefont {Young}\ \emph {et~al.}(2015)\citenamefont {Young},
  \citenamefont {Kapusta}, \citenamefont {Gale}, \citenamefont {Jeon},\ and\
  \citenamefont {Schenke}}]{Young:2014pka}%
  \BibitemOpen
  \bibfield  {author} {\bibinfo {author} {\bibfnamefont {C.}~\bibnamefont
  {Young}}, \bibinfo {author} {\bibfnamefont {J.~I.}\ \bibnamefont {Kapusta}},
  \bibinfo {author} {\bibfnamefont {C.}~\bibnamefont {Gale}}, \bibinfo {author}
  {\bibfnamefont {S.}~\bibnamefont {Jeon}}, \ and\ \bibinfo {author}
  {\bibfnamefont {B.}~\bibnamefont {Schenke}},\ }\href {\doibase
  10.1103/PhysRevC.91.044901} {\bibfield  {journal} {\bibinfo  {journal} {Phys.
  Rev. C}\ }\textbf {\bibinfo {volume} {91}},\ \bibinfo {pages} {044901}
  (\bibinfo {year} {2015})},\ \Eprint {http://arxiv.org/abs/1407.1077}
  {arXiv:1407.1077 [nucl-th]} \BibitemShut {NoStop}%
\bibitem [{\citenamefont {Berges}\ \emph {et~al.}(2010)\citenamefont {Berges},
  \citenamefont {Schlichting},\ and\ \citenamefont {Sexty}}]{Berges:2009jz}%
  \BibitemOpen
  \bibfield  {author} {\bibinfo {author} {\bibfnamefont {J.}~\bibnamefont
  {Berges}}, \bibinfo {author} {\bibfnamefont {S.}~\bibnamefont {Schlichting}},
  \ and\ \bibinfo {author} {\bibfnamefont {D.}~\bibnamefont {Sexty}},\
  }\href@noop {} {\bibfield  {journal} {\bibinfo  {journal} {Nucl. Phys. B}\
  }\textbf {\bibinfo {volume} {832}},\ \bibinfo {pages} {228} (\bibinfo {year}
  {2010})},\ \Eprint {http://arxiv.org/abs/0912.3135} {arXiv:0912.3135
  [hep-lat]} \BibitemShut {NoStop}%
\bibitem [{\citenamefont {Schweitzer}\ \emph {et~al.}(2020)\citenamefont
  {Schweitzer}, \citenamefont {Schlichting},\ and\ \citenamefont {von
  Smekal}}]{Schweitzer:2020noq}%
  \BibitemOpen
  \bibfield  {author} {\bibinfo {author} {\bibfnamefont {D.}~\bibnamefont
  {Schweitzer}}, \bibinfo {author} {\bibfnamefont {S.}~\bibnamefont
  {Schlichting}}, \ and\ \bibinfo {author} {\bibfnamefont {L.}~\bibnamefont
  {von Smekal}},\ }\href@noop {} {\bibfield  {journal} {\bibinfo  {journal}
  {Nucl. Phys. B}\ }\textbf {\bibinfo {volume} {960}},\ \bibinfo {pages}
  {115165} (\bibinfo {year} {2020})},\ \Eprint
  {http://arxiv.org/abs/2007.03374} {arXiv:2007.03374 [hep-lat]} \BibitemShut
  {NoStop}%
\bibitem [{\citenamefont {Schweitzer}\ \emph {et~al.}(2022)\citenamefont
  {Schweitzer}, \citenamefont {Schlichting},\ and\ \citenamefont {von
  Smekal}}]{Schweitzer:2021iqk}%
  \BibitemOpen
  \bibfield  {author} {\bibinfo {author} {\bibfnamefont {D.}~\bibnamefont
  {Schweitzer}}, \bibinfo {author} {\bibfnamefont {S.}~\bibnamefont
  {Schlichting}}, \ and\ \bibinfo {author} {\bibfnamefont {L.}~\bibnamefont
  {von Smekal}},\ }\href {\doibase 10.1016/j.nuclphysb.2022.115944} {\bibfield
  {journal} {\bibinfo  {journal} {Nucl. Phys. B}\ }\textbf {\bibinfo {volume}
  {984}},\ \bibinfo {pages} {115944} (\bibinfo {year} {2022})},\ \Eprint
  {http://arxiv.org/abs/2110.01696} {arXiv:2110.01696 [hep-lat]} \BibitemShut
  {NoStop}%
\bibitem [{\citenamefont {Nahrgang}\ \emph {et~al.}(2019)\citenamefont
  {Nahrgang}, \citenamefont {Bluhm}, \citenamefont {Sch{\"a}fer},\ and\
  \citenamefont {Bass}}]{Nahrgang:2018afz}%
  \BibitemOpen
  \bibfield  {author} {\bibinfo {author} {\bibfnamefont {M.}~\bibnamefont
  {Nahrgang}}, \bibinfo {author} {\bibfnamefont {M.}~\bibnamefont {Bluhm}},
  \bibinfo {author} {\bibfnamefont {T.}~\bibnamefont {Sch{\"a}fer}}, \ and\
  \bibinfo {author} {\bibfnamefont {S.~A.}\ \bibnamefont {Bass}},\ }\href
  {\doibase 10.1103/PhysRevD.99.116015} {\bibfield  {journal} {\bibinfo
  {journal} {Phys. Rev. D}\ }\textbf {\bibinfo {volume} {99}},\ \bibinfo
  {pages} {116015} (\bibinfo {year} {2019})},\ \Eprint
  {http://arxiv.org/abs/1804.05728} {arXiv:1804.05728 [nucl-th]} \BibitemShut
  {NoStop}%
\bibitem [{\citenamefont {Pihan}\ \emph {et~al.}(2023)\citenamefont {Pihan},
  \citenamefont {Bluhm}, \citenamefont {Kitazawa}, \citenamefont {Sami},\ and\
  \citenamefont {Nahrgang}}]{Pihan:2022xcl}%
  \BibitemOpen
  \bibfield  {author} {\bibinfo {author} {\bibfnamefont {G.}~\bibnamefont
  {Pihan}}, \bibinfo {author} {\bibfnamefont {M.}~\bibnamefont {Bluhm}},
  \bibinfo {author} {\bibfnamefont {M.}~\bibnamefont {Kitazawa}}, \bibinfo
  {author} {\bibfnamefont {T.}~\bibnamefont {Sami}}, \ and\ \bibinfo {author}
  {\bibfnamefont {M.}~\bibnamefont {Nahrgang}},\ }\href {\doibase
  10.1103/PhysRevC.107.014908} {\bibfield  {journal} {\bibinfo  {journal}
  {Phys. Rev. C}\ }\textbf {\bibinfo {volume} {107}},\ \bibinfo {pages}
  {014908} (\bibinfo {year} {2023})},\ \Eprint
  {http://arxiv.org/abs/2205.12834} {arXiv:2205.12834 [nucl-th]} \BibitemShut
  {NoStop}%
\bibitem [{\citenamefont {Kuznietsov}\ \emph {et~al.}(2022)\citenamefont
  {Kuznietsov}, \citenamefont {Savchuk}, \citenamefont {Gorenstein},
  \citenamefont {Koch},\ and\ \citenamefont {Vovchenko}}]{Kuznietsov:2022pcn}%
  \BibitemOpen
  \bibfield  {author} {\bibinfo {author} {\bibfnamefont {V.~A.}\ \bibnamefont
  {Kuznietsov}}, \bibinfo {author} {\bibfnamefont {O.}~\bibnamefont {Savchuk}},
  \bibinfo {author} {\bibfnamefont {M.~I.}\ \bibnamefont {Gorenstein}},
  \bibinfo {author} {\bibfnamefont {V.}~\bibnamefont {Koch}}, \ and\ \bibinfo
  {author} {\bibfnamefont {V.}~\bibnamefont {Vovchenko}},\ }\href {\doibase
  10.1103/PhysRevC.105.044903} {\bibfield  {journal} {\bibinfo  {journal}
  {Phys. Rev. C}\ }\textbf {\bibinfo {volume} {105}},\ \bibinfo {pages}
  {044903} (\bibinfo {year} {2022})},\ \Eprint
  {http://arxiv.org/abs/2201.08486} {arXiv:2201.08486 [hep-ph]} \BibitemShut
  {NoStop}%
\bibitem [{\citenamefont {Chen}\ \emph {et~al.}(2005)\citenamefont {Chen},
  \citenamefont {Chimowitz}, \citenamefont {De},\ and\ \citenamefont
  {Shapir}}]{Chen:1995}%
  \BibitemOpen
  \bibfield  {author} {\bibinfo {author} {\bibfnamefont {A.}~\bibnamefont
  {Chen}}, \bibinfo {author} {\bibfnamefont {E.~H.}\ \bibnamefont {Chimowitz}},
  \bibinfo {author} {\bibfnamefont {S.}~\bibnamefont {De}}, \ and\ \bibinfo
  {author} {\bibfnamefont {Y.}~\bibnamefont {Shapir}},\ }\href {\doibase
  10.1103/PhysRevLett.95.255701} {\bibfield  {journal} {\bibinfo  {journal}
  {Phys. Rev. Lett.}\ }\textbf {\bibinfo {volume} {95}},\ \bibinfo {pages}
  {255701} (\bibinfo {year} {2005})}\BibitemShut {NoStop}%
\bibitem [{\citenamefont {Florio}\ \emph {et~al.}(2022)\citenamefont {Florio},
  \citenamefont {Grossi}, \citenamefont {Soloviev},\ and\ \citenamefont
  {Teaney}}]{Florio:2021jlx}%
  \BibitemOpen
  \bibfield  {author} {\bibinfo {author} {\bibfnamefont {A.}~\bibnamefont
  {Florio}}, \bibinfo {author} {\bibfnamefont {E.}~\bibnamefont {Grossi}},
  \bibinfo {author} {\bibfnamefont {A.}~\bibnamefont {Soloviev}}, \ and\
  \bibinfo {author} {\bibfnamefont {D.}~\bibnamefont {Teaney}},\ }\href@noop {}
  {\bibfield  {journal} {\bibinfo  {journal} {Phys. Rev. D}\ }\textbf {\bibinfo
  {volume} {105}},\ \bibinfo {pages} {054512} (\bibinfo {year} {2022})},\
  \Eprint {http://arxiv.org/abs/2111.03640} {arXiv:2111.03640 [hep-lat]}
  \BibitemShut {NoStop}%
\bibitem [{\citenamefont {Sch{\"a}fer}\ and\ \citenamefont
  {Skokov}(2022)}]{Schaefer:2022bfm}%
  \BibitemOpen
  \bibfield  {author} {\bibinfo {author} {\bibfnamefont {T.}~\bibnamefont
  {Sch{\"a}fer}}\ and\ \bibinfo {author} {\bibfnamefont {V.}~\bibnamefont
  {Skokov}},\ }\href@noop {} {\bibfield  {journal} {\bibinfo  {journal} {Phys.
  Rev. D}\ }\textbf {\bibinfo {volume} {106}},\ \bibinfo {pages} {014006}
  (\bibinfo {year} {2022})},\ \Eprint {http://arxiv.org/abs/2204.02433}
  {arXiv:2204.02433 [nucl-th]} \BibitemShut {NoStop}%
\bibitem [{\citenamefont {Florio}\ \emph {et~al.}(2024)\citenamefont {Florio},
  \citenamefont {Grossi},\ and\ \citenamefont {Teaney}}]{Florio:2023kmy}%
  \BibitemOpen
  \bibfield  {author} {\bibinfo {author} {\bibfnamefont {A.}~\bibnamefont
  {Florio}}, \bibinfo {author} {\bibfnamefont {E.}~\bibnamefont {Grossi}}, \
  and\ \bibinfo {author} {\bibfnamefont {D.}~\bibnamefont {Teaney}},\
  }\href@noop {} {\bibfield  {journal} {\bibinfo  {journal} {Phys. Rev. D}\
  }\textbf {\bibinfo {volume} {109}},\ \bibinfo {pages} {054037} (\bibinfo
  {year} {2024})},\ \Eprint {http://arxiv.org/abs/2306.06887}
  {arXiv:2306.06887} \BibitemShut {NoStop}%
\bibitem [{\citenamefont {Chattopadhyay}\ \emph {et~al.}(2023)\citenamefont
  {Chattopadhyay}, \citenamefont {Ott}, \citenamefont {Sch{\"a}fer},\ and\
  \citenamefont {Skokov}}]{Chattopadhyay:2023jfm}%
  \BibitemOpen
  \bibfield  {author} {\bibinfo {author} {\bibfnamefont {C.}~\bibnamefont
  {Chattopadhyay}}, \bibinfo {author} {\bibfnamefont {J.}~\bibnamefont {Ott}},
  \bibinfo {author} {\bibfnamefont {T.}~\bibnamefont {Sch{\"a}fer}}, \ and\
  \bibinfo {author} {\bibfnamefont {V.}~\bibnamefont {Skokov}},\ }\href@noop {}
  {\bibfield  {journal} {\bibinfo  {journal} {Phys. Rev. D}\ }\textbf {\bibinfo
  {volume} {108}},\ \bibinfo {pages} {074004} (\bibinfo {year} {2023})},\
  \Eprint {http://arxiv.org/abs/2304.07279} {arXiv:2304.07279 [nucl-th]}
  \BibitemShut {NoStop}%
\bibitem [{\citenamefont {Ba\c{s}ar}\ \emph {et~al.}(2024)\citenamefont
  {Ba\c{s}ar}, \citenamefont {Bhambure}, \citenamefont {Singh},\ and\
  \citenamefont {Teaney}}]{Basar:2024qxd}%
  \BibitemOpen
  \bibfield  {author} {\bibinfo {author} {\bibfnamefont {G.}~\bibnamefont
  {Ba\c{s}ar}}, \bibinfo {author} {\bibfnamefont {J.}~\bibnamefont {Bhambure}},
  \bibinfo {author} {\bibfnamefont {R.}~\bibnamefont {Singh}}, \ and\ \bibinfo
  {author} {\bibfnamefont {D.}~\bibnamefont {Teaney}},\ }\href {\doibase
  10.1103/PhysRevC.110.044903} {\bibfield  {journal} {\bibinfo  {journal}
  {Phys. Rev. C}\ }\textbf {\bibinfo {volume} {110}},\ \bibinfo {pages}
  {044903} (\bibinfo {year} {2024})},\ \Eprint
  {http://arxiv.org/abs/2403.04185} {arXiv:2403.04185 [nucl-th]} \BibitemShut
  {NoStop}%
\bibitem [{\citenamefont {Chattopadhyay}\ \emph {et~al.}(2024)\citenamefont
  {Chattopadhyay}, \citenamefont {Ott}, \citenamefont {Schaefer},\ and\
  \citenamefont {Skokov}}]{Chattopadhyay:2024jlh}%
  \BibitemOpen
  \bibfield  {author} {\bibinfo {author} {\bibfnamefont {C.}~\bibnamefont
  {Chattopadhyay}}, \bibinfo {author} {\bibfnamefont {J.}~\bibnamefont {Ott}},
  \bibinfo {author} {\bibfnamefont {T.}~\bibnamefont {Schaefer}}, \ and\
  \bibinfo {author} {\bibfnamefont {V.~V.}\ \bibnamefont {Skokov}},\ }\href
  {\doibase 10.1103/PhysRevLett.133.032301} {\bibfield  {journal} {\bibinfo
  {journal} {Phys. Rev. Lett.}\ }\textbf {\bibinfo {volume} {133}},\ \bibinfo
  {pages} {032301} (\bibinfo {year} {2024})},\ \Eprint
  {http://arxiv.org/abs/2403.10608} {arXiv:2403.10608 [nucl-th]} \BibitemShut
  {NoStop}%
\bibitem [{\citenamefont {Gao}\ \emph {et~al.}(2021)\citenamefont {Gao},
  \citenamefont {Kirkpatrick}, \citenamefont {Marzuola}, \citenamefont
  {Mattingly},\ and\ \citenamefont {Newhall}}]{Gao:2021}%
  \BibitemOpen
  \bibfield  {author} {\bibinfo {author} {\bibfnamefont {Y.}~\bibnamefont
  {Gao}}, \bibinfo {author} {\bibfnamefont {K.}~\bibnamefont {Kirkpatrick}},
  \bibinfo {author} {\bibfnamefont {J.}~\bibnamefont {Marzuola}}, \bibinfo
  {author} {\bibfnamefont {J.}~\bibnamefont {Mattingly}}, \ and\ \bibinfo
  {author} {\bibfnamefont {K.~A.}\ \bibnamefont {Newhall}},\ }\href@noop {}
  {\bibfield  {journal} {\bibinfo  {journal} {Communications in Mathematical
  Sciences}\ }\textbf {\bibinfo {volume} {19}},\ \bibinfo {pages} {453}
  (\bibinfo {year} {2021})},\ \Eprint {http://arxiv.org/abs/1806.05282}
  {arXiv:1806.05282 [math.PR]} \BibitemShut {NoStop}%
\bibitem [{\citenamefont {Folk}\ and\ \citenamefont
  {Moser}(2006)}]{Folk:2006ve}%
  \BibitemOpen
  \bibfield  {author} {\bibinfo {author} {\bibfnamefont {R.}~\bibnamefont
  {Folk}}\ and\ \bibinfo {author} {\bibfnamefont {H.-G.}\ \bibnamefont
  {Moser}},\ }\href {\doibase 10.1088/0305-4470/39/24/R01} {\bibfield
  {journal} {\bibinfo  {journal} {J. Phys. A}\ }\textbf {\bibinfo {volume}
  {39}},\ \bibinfo {pages} {R207} (\bibinfo {year} {2006})}\BibitemShut
  {NoStop}%
\bibitem [{\citenamefont {An}\ \emph {et~al.}(2023)\citenamefont {An},
  \citenamefont {Basar}, \citenamefont {Stephanov},\ and\ \citenamefont
  {Yee}}]{An:2022jgc}%
  \BibitemOpen
  \bibfield  {author} {\bibinfo {author} {\bibfnamefont {X.}~\bibnamefont
  {An}}, \bibinfo {author} {\bibfnamefont {G.}~\bibnamefont {Basar}}, \bibinfo
  {author} {\bibfnamefont {M.}~\bibnamefont {Stephanov}}, \ and\ \bibinfo
  {author} {\bibfnamefont {H.-U.}\ \bibnamefont {Yee}},\ }\href {\doibase
  10.1103/PhysRevC.108.034910} {\bibfield  {journal} {\bibinfo  {journal}
  {Phys. Rev. C}\ }\textbf {\bibinfo {volume} {108}},\ \bibinfo {pages}
  {034910} (\bibinfo {year} {2023})},\ \Eprint
  {http://arxiv.org/abs/2212.14029} {arXiv:2212.14029 [hep-th]} \BibitemShut
  {NoStop}%
\bibitem [{\citenamefont {Parotto}\ \emph {et~al.}(2020)\citenamefont
  {Parotto}, \citenamefont {Bluhm}, \citenamefont {Mroczek}, \citenamefont
  {Nahrgang}, \citenamefont {Noronha-Hostler}, \citenamefont {Rajagopal},
  \citenamefont {Ratti}, \citenamefont {Sch\"afer},\ and\ \citenamefont
  {Stephanov}}]{Parotto:2018pwx}%
  \BibitemOpen
  \bibfield  {author} {\bibinfo {author} {\bibfnamefont {P.}~\bibnamefont
  {Parotto}}, \bibinfo {author} {\bibfnamefont {M.}~\bibnamefont {Bluhm}},
  \bibinfo {author} {\bibfnamefont {D.}~\bibnamefont {Mroczek}}, \bibinfo
  {author} {\bibfnamefont {M.}~\bibnamefont {Nahrgang}}, \bibinfo {author}
  {\bibfnamefont {J.}~\bibnamefont {Noronha-Hostler}}, \bibinfo {author}
  {\bibfnamefont {K.}~\bibnamefont {Rajagopal}}, \bibinfo {author}
  {\bibfnamefont {C.}~\bibnamefont {Ratti}}, \bibinfo {author} {\bibfnamefont
  {T.}~\bibnamefont {Sch\"afer}}, \ and\ \bibinfo {author} {\bibfnamefont
  {M.}~\bibnamefont {Stephanov}},\ }\href {\doibase
  10.1103/PhysRevC.101.034901} {\bibfield  {journal} {\bibinfo  {journal}
  {Phys. Rev. C}\ }\textbf {\bibinfo {volume} {101}},\ \bibinfo {pages}
  {034901} (\bibinfo {year} {2020})},\ \Eprint
  {http://arxiv.org/abs/1805.05249} {arXiv:1805.05249 [hep-ph]} \BibitemShut
  {NoStop}%
\bibitem [{\citenamefont {Kahangirwe}\ \emph {et~al.}(2024)\citenamefont
  {Kahangirwe}, \citenamefont {Bass}, \citenamefont {Bratkovskaya},
  \citenamefont {Jahan}, \citenamefont {Moreau}, \citenamefont {Parotto},
  \citenamefont {Price}, \citenamefont {Ratti}, \citenamefont {Soloveva},\ and\
  \citenamefont {Stephanov}}]{Kahangirwe:2024cny}%
  \BibitemOpen
  \bibfield  {author} {\bibinfo {author} {\bibfnamefont {M.}~\bibnamefont
  {Kahangirwe}}, \bibinfo {author} {\bibfnamefont {S.~A.}\ \bibnamefont
  {Bass}}, \bibinfo {author} {\bibfnamefont {E.}~\bibnamefont {Bratkovskaya}},
  \bibinfo {author} {\bibfnamefont {J.}~\bibnamefont {Jahan}}, \bibinfo
  {author} {\bibfnamefont {P.}~\bibnamefont {Moreau}}, \bibinfo {author}
  {\bibfnamefont {P.}~\bibnamefont {Parotto}}, \bibinfo {author} {\bibfnamefont
  {D.}~\bibnamefont {Price}}, \bibinfo {author} {\bibfnamefont
  {C.}~\bibnamefont {Ratti}}, \bibinfo {author} {\bibfnamefont
  {O.}~\bibnamefont {Soloveva}}, \ and\ \bibinfo {author} {\bibfnamefont
  {M.}~\bibnamefont {Stephanov}},\ }\href {\doibase
  10.1103/PhysRevD.109.094046} {\bibfield  {journal} {\bibinfo  {journal}
  {Phys. Rev. D}\ }\textbf {\bibinfo {volume} {109}},\ \bibinfo {pages}
  {094046} (\bibinfo {year} {2024})},\ \Eprint
  {http://arxiv.org/abs/2402.08636} {arXiv:2402.08636 [nucl-th]} \BibitemShut
  {NoStop}%
\bibitem [{\citenamefont {Onuki}(1997)}]{Onuki:1997}%
  \BibitemOpen
  \bibfield  {author} {\bibinfo {author} {\bibfnamefont {A.}~\bibnamefont
  {Onuki}},\ }\href {\doibase 10.1103/PhysRevE.55.403} {\bibfield  {journal}
  {\bibinfo  {journal} {Phys. Rev. E}\ }\textbf {\bibinfo {volume} {55}},\
  \bibinfo {pages} {403} (\bibinfo {year} {1997})}\BibitemShut {NoStop}%
\bibitem [{\citenamefont {Martinez}\ \emph {et~al.}(2019)\citenamefont
  {Martinez}, \citenamefont {Sch\"afer},\ and\ \citenamefont
  {Skokov}}]{Martinez:2019bsn}%
  \BibitemOpen
  \bibfield  {author} {\bibinfo {author} {\bibfnamefont {M.}~\bibnamefont
  {Martinez}}, \bibinfo {author} {\bibfnamefont {T.}~\bibnamefont {Sch\"afer}},
  \ and\ \bibinfo {author} {\bibfnamefont {V.}~\bibnamefont {Skokov}},\ }\href
  {\doibase 10.1103/PhysRevD.100.074017} {\bibfield  {journal} {\bibinfo
  {journal} {Phys. Rev. D}\ }\textbf {\bibinfo {volume} {100}},\ \bibinfo
  {pages} {074017} (\bibinfo {year} {2019})},\ \Eprint
  {http://arxiv.org/abs/1906.11306} {arXiv:1906.11306 [hep-ph]} \BibitemShut
  {NoStop}%
\bibitem [{\citenamefont {{Onuki}}(2002)}]{Onuki:2002}%
  \BibitemOpen
  \bibfield  {author} {\bibinfo {author} {\bibfnamefont {A.}~\bibnamefont
  {{Onuki}}},\ }\href@noop {} {\emph {\bibinfo {title} {{Phase Transition
  Dynamics}}}}\ (\bibinfo  {publisher} {Cambridge University Press},\ \bibinfo
  {year} {2002})\BibitemShut {NoStop}%
\bibitem [{\citenamefont {{Vasil'ev}}(2004)}]{Vasilev:2004}%
  \BibitemOpen
  \bibfield  {author} {\bibinfo {author} {\bibfnamefont {A.}~\bibnamefont
  {{Vasil'ev}}},\ }\href@noop {} {\emph {\bibinfo {title} {{The Field Theoretic
  Renormalization Group in Critical Behavior Theory and Stochastic
  Dynamics}}}}\ (\bibinfo  {publisher} {Chapman \& Hall/CRC},\ \bibinfo {year}
  {2004})\BibitemShut {NoStop}%
\bibitem [{\citenamefont {Antonov}\ and\ \citenamefont
  {Vasil'ev}(1984)}]{Antonov:1984}%
  \BibitemOpen
  \bibfield  {author} {\bibinfo {author} {\bibfnamefont {N.~V.}\ \bibnamefont
  {Antonov}}\ and\ \bibinfo {author} {\bibfnamefont {A.~N.}\ \bibnamefont
  {Vasil'ev}},\ }\href@noop {} {\bibfield  {journal} {\bibinfo  {journal}
  {Theor. Math. Phys.}\ }\textbf {\bibinfo {volume} {60}},\ \bibinfo {pages}
  {671} (\bibinfo {year} {1984})}\BibitemShut {NoStop}%
\bibitem [{\citenamefont {Folk}\ and\ \citenamefont {Moser}(1998)}]{Folk:1998}%
  \BibitemOpen
  \bibfield  {author} {\bibinfo {author} {\bibfnamefont {R.}~\bibnamefont
  {Folk}}\ and\ \bibinfo {author} {\bibfnamefont {G.}~\bibnamefont {Moser}},\
  }\href@noop {} {\bibfield  {journal} {\bibinfo  {journal} {Phys. Rev. E}\
  }\textbf {\bibinfo {volume} {57}},\ \bibinfo {pages} {683} (\bibinfo {year}
  {1998})}\BibitemShut {NoStop}%
\bibitem [{\citenamefont {{Dzyaloshinskii}}\ and\ \citenamefont
  {{Volovick}}(1980)}]{Dzyaloshinskii:1980}%
  \BibitemOpen
  \bibfield  {author} {\bibinfo {author} {\bibfnamefont {I.~E.}\ \bibnamefont
  {{Dzyaloshinskii}}}\ and\ \bibinfo {author} {\bibfnamefont {G.~E.}\
  \bibnamefont {{Volovick}}},\ }\href@noop {} {\bibfield  {journal} {\bibinfo
  {journal} {Annals of Physics}\ }\textbf {\bibinfo {volume} {125}},\ \bibinfo
  {pages} {67} (\bibinfo {year} {1980})}\BibitemShut {NoStop}%
\bibitem [{\citenamefont {Morinishi}\ \emph {et~al.}(1998)\citenamefont
  {Morinishi}, \citenamefont {Lund}, \citenamefont {Vasilyev},\ and\
  \citenamefont {Moin}}]{Morinishi:1998}%
  \BibitemOpen
  \bibfield  {author} {\bibinfo {author} {\bibfnamefont {Y.}~\bibnamefont
  {Morinishi}}, \bibinfo {author} {\bibfnamefont {T.}~\bibnamefont {Lund}},
  \bibinfo {author} {\bibfnamefont {O.}~\bibnamefont {Vasilyev}}, \ and\
  \bibinfo {author} {\bibfnamefont {P.}~\bibnamefont {Moin}},\ }\href@noop {}
  {\bibfield  {journal} {\bibinfo  {journal} {Journal of Computational
  Physics}\ }\textbf {\bibinfo {volume} {143}},\ \bibinfo {pages} {90}
  (\bibinfo {year} {1998})}\BibitemShut {NoStop}%
\bibitem [{\citenamefont {Shu}\ and\ \citenamefont {Osher}(1988)}]{Shu:1988}%
  \BibitemOpen
  \bibfield  {author} {\bibinfo {author} {\bibfnamefont {C.-W.}\ \bibnamefont
  {Shu}}\ and\ \bibinfo {author} {\bibfnamefont {S.}~\bibnamefont {Osher}},\
  }\href@noop {} {\bibfield  {journal} {\bibinfo  {journal} {Journal of
  Computational Physics}\ }\textbf {\bibinfo {volume} {77}},\ \bibinfo {pages}
  {439} (\bibinfo {year} {1988})}\BibitemShut {NoStop}%
\bibitem [{\citenamefont {Alday}\ and\ \citenamefont
  {Zhiboedov}(2016)}]{Alday:2015ota}%
  \BibitemOpen
  \bibfield  {author} {\bibinfo {author} {\bibfnamefont {L.~F.}\ \bibnamefont
  {Alday}}\ and\ \bibinfo {author} {\bibfnamefont {A.}~\bibnamefont
  {Zhiboedov}},\ }\href {\doibase 10.1007/JHEP06(2016)091} {\bibfield
  {journal} {\bibinfo  {journal} {JHEP}\ }\textbf {\bibinfo {volume} {06}},\
  \bibinfo {pages} {091} (\bibinfo {year} {2016})},\ \Eprint
  {http://arxiv.org/abs/1506.04659} {arXiv:1506.04659 [hep-th]} \BibitemShut
  {NoStop}%
\bibitem [{\citenamefont {El-Showk}\ \emph {et~al.}(2014)\citenamefont
  {El-Showk}, \citenamefont {Paulos}, \citenamefont {Poland}, \citenamefont
  {Rychkov}, \citenamefont {Simmons-Duffin},\ and\ \citenamefont
  {Vichi}}]{El-Showk:2014dwa}%
  \BibitemOpen
  \bibfield  {author} {\bibinfo {author} {\bibfnamefont {S.}~\bibnamefont
  {El-Showk}}, \bibinfo {author} {\bibfnamefont {M.~F.}\ \bibnamefont
  {Paulos}}, \bibinfo {author} {\bibfnamefont {D.}~\bibnamefont {Poland}},
  \bibinfo {author} {\bibfnamefont {S.}~\bibnamefont {Rychkov}}, \bibinfo
  {author} {\bibfnamefont {D.}~\bibnamefont {Simmons-Duffin}}, \ and\ \bibinfo
  {author} {\bibfnamefont {A.}~\bibnamefont {Vichi}},\ }\href {\doibase
  10.1007/s10955-014-1042-7} {\bibfield  {journal} {\bibinfo  {journal} {J.
  Stat. Phys.}\ }\textbf {\bibinfo {volume} {157}},\ \bibinfo {pages} {869}
  (\bibinfo {year} {2014})},\ \Eprint {http://arxiv.org/abs/1403.4545}
  {arXiv:1403.4545 [hep-th]} \BibitemShut {NoStop}%
\bibitem [{\citenamefont {Chafin}\ and\ \citenamefont
  {Sch\"afer}(2013)}]{Chafin:2012eq}%
  \BibitemOpen
  \bibfield  {author} {\bibinfo {author} {\bibfnamefont {C.}~\bibnamefont
  {Chafin}}\ and\ \bibinfo {author} {\bibfnamefont {T.}~\bibnamefont
  {Sch\"afer}},\ }\href@noop {} {\bibfield  {journal} {\bibinfo  {journal}
  {Phys. Rev. A}\ }\textbf {\bibinfo {volume} {87}},\ \bibinfo {pages} {023629}
  (\bibinfo {year} {2013})},\ \Eprint {http://arxiv.org/abs/1209.1006}
  {arXiv:1209.1006 [cond-mat.quant-gas]} \BibitemShut {NoStop}%
\bibitem [{\citenamefont {Kawasaki}(1970)}]{Kawasaki:1970}%
  \BibitemOpen
  \bibfield  {author} {\bibinfo {author} {\bibfnamefont {K.}~\bibnamefont
  {Kawasaki}},\ }\href@noop {} {\bibfield  {journal} {\bibinfo  {journal}
  {Annals of Physics}\ }\textbf {\bibinfo {volume} {61}},\ \bibinfo {pages} {1}
  (\bibinfo {year} {1970})}\BibitemShut {NoStop}%
\bibitem [{\citenamefont {Adzhemyan}\ \emph {et~al.}(1999)\citenamefont
  {Adzhemyan}, \citenamefont {Vasiliev}, \citenamefont {Kabrits},\ and\
  \citenamefont {Kompaniets}}]{Adzhemyan:1999h}%
  \BibitemOpen
  \bibfield  {author} {\bibinfo {author} {\bibfnamefont {L.~T.}\ \bibnamefont
  {Adzhemyan}}, \bibinfo {author} {\bibfnamefont {A.}~\bibnamefont {Vasiliev}},
  \bibinfo {author} {\bibfnamefont {Y.~S.}\ \bibnamefont {Kabrits}}, \ and\
  \bibinfo {author} {\bibfnamefont {M.~V.}\ \bibnamefont {Kompaniets}},\
  }\href@noop {} {\bibfield  {journal} {\bibinfo  {journal} {Theoretical and
  Mathematical Physics}\ }\textbf {\bibinfo {volume} {119}},\ \bibinfo {pages}
  {454} (\bibinfo {year} {1999})}\BibitemShut {NoStop}%
\bibitem [{\citenamefont {{Binder}}(1981)}]{Binder:1981}%
  \BibitemOpen
  \bibfield  {author} {\bibinfo {author} {\bibfnamefont {K.}~\bibnamefont
  {{Binder}}},\ }\href {\doibase 10.1007/BF01293604} {\bibfield  {journal}
  {\bibinfo  {journal} {Zeitschrift fur Physik B Condensed Matter}\ }\textbf
  {\bibinfo {volume} {43}},\ \bibinfo {pages} {119} (\bibinfo {year}
  {1981})}\BibitemShut {NoStop}%
\bibitem [{\citenamefont {Hasenbusch}\ \emph {et~al.}(1998)\citenamefont
  {Hasenbusch}, \citenamefont {Pinn},\ and\ \citenamefont
  {Vinti}}]{Hasenbusch:1998ve}%
  \BibitemOpen
  \bibfield  {author} {\bibinfo {author} {\bibfnamefont {M.}~\bibnamefont
  {Hasenbusch}}, \bibinfo {author} {\bibfnamefont {K.}~\bibnamefont {Pinn}}, \
  and\ \bibinfo {author} {\bibfnamefont {S.}~\bibnamefont {Vinti}},\
  }\href@noop {} {\  (\bibinfo {year} {1998})},\ \Eprint
  {http://arxiv.org/abs/cond-mat/9804186} {arXiv:cond-mat/9804186} \BibitemShut
  {NoStop}%
\bibitem [{\citenamefont {Kamieniarz}\ and\ \citenamefont
  {Blote}(1993)}]{Kamieniarz:1993}%
  \BibitemOpen
  \bibfield  {author} {\bibinfo {author} {\bibfnamefont {G.}~\bibnamefont
  {Kamieniarz}}\ and\ \bibinfo {author} {\bibfnamefont {H.~W.~J.}\ \bibnamefont
  {Blote}},\ }\href@noop {} {\bibfield  {journal} {\bibinfo  {journal} {Journal
  of Physics A: Mathematical and General}\ }\textbf {\bibinfo {volume} {26}},\
  \bibinfo {pages} {201} (\bibinfo {year} {1993})}\BibitemShut {NoStop}%
\bibitem [{\citenamefont {Francesco}\ \emph {et~al.}(2012)\citenamefont
  {Francesco}, \citenamefont {Mathieu},\ and\ \citenamefont
  {Senechal}}]{Francesco:2012}%
  \BibitemOpen
  \bibfield  {author} {\bibinfo {author} {\bibfnamefont {P.}~\bibnamefont
  {Francesco}}, \bibinfo {author} {\bibfnamefont {P.}~\bibnamefont {Mathieu}},
  \ and\ \bibinfo {author} {\bibfnamefont {D.}~\bibnamefont {Senechal}},\
  }\href {https://books.google.com/books?id=5u7jBwAAQBAJ} {\emph {\bibinfo
  {title} {Conformal Field Theory}}},\ Graduate Texts in Contemporary Physics\
  (\bibinfo  {publisher} {Springer New York},\ \bibinfo {year}
  {2012})\BibitemShut {NoStop}%
\bibitem [{\citenamefont {Ohta}(1975)}]{Ohta:1975}%
  \BibitemOpen
  \bibfield  {author} {\bibinfo {author} {\bibfnamefont {T.}~\bibnamefont
  {Ohta}},\ }\href {\doibase 10.1143/PTP.54.1566} {\bibfield  {journal}
  {\bibinfo  {journal} {Progress of Theoretical Physics}\ }\textbf {\bibinfo
  {volume} {54}},\ \bibinfo {pages} {1566} (\bibinfo {year}
  {1975})}\BibitemShut {NoStop}%
\bibitem [{\citenamefont {Kadanoff}\ \emph {et~al.}(1989)\citenamefont
  {Kadanoff}, \citenamefont {McNamara},\ and\ \citenamefont
  {Zanetti}}]{Kadanoff:1989}%
  \BibitemOpen
  \bibfield  {author} {\bibinfo {author} {\bibfnamefont {L.~P.}\ \bibnamefont
  {Kadanoff}}, \bibinfo {author} {\bibfnamefont {G.~R.}\ \bibnamefont
  {McNamara}}, \ and\ \bibinfo {author} {\bibfnamefont {G.}~\bibnamefont
  {Zanetti}},\ }\href@noop {} {\bibfield  {journal} {\bibinfo  {journal} {Phys.
  Rev. A}\ }\textbf {\bibinfo {volume} {40}},\ \bibinfo {pages} {4527}
  (\bibinfo {year} {1989})}\BibitemShut {NoStop}%
\bibitem [{\citenamefont {Kovtun}(2012)}]{Kovtun:2012rj}%
  \BibitemOpen
  \bibfield  {author} {\bibinfo {author} {\bibfnamefont {P.}~\bibnamefont
  {Kovtun}},\ }\href@noop {} {\bibfield  {journal} {\bibinfo  {journal} {J.
  Phys. A}\ }\textbf {\bibinfo {volume} {45}},\ \bibinfo {pages} {473001}
  (\bibinfo {year} {2012})},\ \Eprint {http://arxiv.org/abs/1205.5040}
  {arXiv:1205.5040 [hep-th]} \BibitemShut {NoStop}%
\bibitem [{\citenamefont {Armas}\ and\ \citenamefont
  {Jain}(2021)}]{Armas:2020mpr}%
  \BibitemOpen
  \bibfield  {author} {\bibinfo {author} {\bibfnamefont {J.}~\bibnamefont
  {Armas}}\ and\ \bibinfo {author} {\bibfnamefont {A.}~\bibnamefont {Jain}},\
  }\href {\doibase 10.21468/SciPostPhys.11.3.054} {\bibfield  {journal}
  {\bibinfo  {journal} {SciPost Phys.}\ }\textbf {\bibinfo {volume} {11}},\
  \bibinfo {pages} {054} (\bibinfo {year} {2021})},\ \Eprint
  {http://arxiv.org/abs/2010.15782} {arXiv:2010.15782 [hep-th]} \BibitemShut
  {NoStop}%
\bibitem [{\citenamefont {Bhambure}\ \emph {et~al.}(2024)\citenamefont
  {Bhambure}, \citenamefont {Singh},\ and\ \citenamefont
  {Teaney}}]{Bhambure:2024gnf}%
  \BibitemOpen
  \bibfield  {author} {\bibinfo {author} {\bibfnamefont {J.}~\bibnamefont
  {Bhambure}}, \bibinfo {author} {\bibfnamefont {R.}~\bibnamefont {Singh}}, \
  and\ \bibinfo {author} {\bibfnamefont {D.}~\bibnamefont {Teaney}},\
  }\href@noop {} {\  (\bibinfo {year} {2024})},\ \Eprint
  {http://arxiv.org/abs/2412.10306} {arXiv:2412.10306 [nucl-th]} \BibitemShut
  {NoStop}%
\bibitem [{\citenamefont {Ott}\ \emph {et~al.}(2025{\natexlab{a}})\citenamefont
  {Ott}, \citenamefont {Chattopadhyay}, \citenamefont {Schaefer},\ and\
  \citenamefont {Skokov}}]{Ott:2025a}%
  \BibitemOpen
  \bibfield  {author} {\bibinfo {author} {\bibfnamefont {J.}~\bibnamefont
  {Ott}}, \bibinfo {author} {\bibfnamefont {C.}~\bibnamefont {Chattopadhyay}},
  \bibinfo {author} {\bibfnamefont {T.}~\bibnamefont {Schaefer}}, \ and\
  \bibinfo {author} {\bibfnamefont {V.}~\bibnamefont {Skokov}},\ }\href
  {https://doi.org/10.5281/zenodo.14706997} {} (\bibinfo {year}
  {2025}{\natexlab{a}}),\ \bibinfo {note} {{Z}enodo Model H 2D (v1.0),
  doi:10.5281$/$zenodo.14706997.}\BibitemShut {Stop}%
\bibitem [{\citenamefont {Ott}\ \emph {et~al.}(2025{\natexlab{b}})\citenamefont
  {Ott}, \citenamefont {Chattopadhyay}, \citenamefont {Schaefer},\ and\
  \citenamefont {Skokov}}]{Ott:2025b}%
  \BibitemOpen
  \bibfield  {author} {\bibinfo {author} {\bibfnamefont {J.}~\bibnamefont
  {Ott}}, \bibinfo {author} {\bibfnamefont {C.}~\bibnamefont {Chattopadhyay}},
  \bibinfo {author} {\bibfnamefont {T.}~\bibnamefont {Schaefer}}, \ and\
  \bibinfo {author} {\bibfnamefont {V.}~\bibnamefont {Skokov}},\ }\href
  {https://doi.org/10.5281/zenodo.14707011} {} (\bibinfo {year}
  {2025}{\natexlab{b}}),\ \bibinfo {note} {{Z}enodo Model H 3D (v1.0),
  doi:10.5281$/$zenodo.14706997.}\BibitemShut {Stop}%
\end{thebibliography}%
\end{document}